\documentclass[twocolumn,twocolappendix]{aastex631}

\usepackage{amsmath,amsopn,amsxtra,txfonts}
\usepackage{comment}
\usepackage{natbib}
\usepackage{graphicx}
\usepackage{gensymb}
\usepackage{float}

\newcommand{\kms}{\,km\,s$^{-1}$}
\newcommand{\cm}{cm$^{-2}$}
\newcommand{\vlsr}{$v_\mathrm{LSR}$}

\newcommand{\delv}{$\Delta v$}

\newcommand{\nsl}{16} 
\newcommand{\ncomp}{800} 

\defcitealias{reid2019}{R19}

\accepted{May 5, 2025}

\submitjournal{ApJ}

\shorttitle{UV Properties of Multi-phase Gas Toward the Inner Galaxy}
\shortauthors{Cashman, Fox, Som et al. 2025}

\begin{document}

\title{Ultraviolet Properties of Multi-phase Gas Toward the Inner Galaxy}

\correspondingauthor{Frances Cashman}
\email{fcashman@presby.edu, afox@stsci.edu}

\author[0000-0003-4237-3553]{Frances H. Cashman}
\affiliation{Department of Physics,
Presbyterian College,
Clinton, SC 29325, USA}
\affiliation{Space Telescope Science Institute, 
3700 San Martin Drive,
Baltimore, MD 21218, USA}

\author[0000-0003-0724-4115]{Andrew J. Fox}
\affil{AURA for ESA, Space Telescope Science Institute, 
3700 San Martin Drive, Baltimore, MD 21218}

\author[0000-0002-4814-2492]{Debopam Som}
\affiliation{Space Telescope Science Institute, 
3700 San Martin Drive,
Baltimore, MD 21218, USA}

\author[0000-0002-0507-7096]{Bart P. Wakker}
\affiliation{Eureka Scientific, 
2452 Delmer Street, Suite 100, 
Oakland, CA 96402, USA}

\author[0000-0002-8109-2642]{Robert A. Benjamin}
\affiliation{Department of Physics, University of Wisconsin-Whitewater, 
800 West Main Street, 
Whitewater, WI 53190, USA}

\author[0000-0002-7955-7359]{Dhanesh Krishnarao}
\affiliation{Department of Physics, Colorado College, 
14 East Cache La Poudre Street, 
Colorado Springs, CO 80903, USA}

\author[0000-0003-3681-0016]{David M. French}
\affiliation{Space Telescope Science Institute, 
3700 San Martin Drive, Baltimore, 
MD 21218, USA}

\author[0000-0002-3120-7173]{Rongmon Bordoloi}
\affiliation{Department of Physics, North Carolina State University, 
421 Riddick Hall, 
Raleigh, NC 27695-8202, USA}

\author[0000-0002-6050-2008]{Felix J. Lockman}
\affiliation{Green Bank Observatory, P.O. Box 2, Rt. 28/92, Green Bank, WV 24944, USA}

\begin{abstract}
We present a systematic study of the multiphase interstellar gas in the Inner Galaxy using HST/STIS absorption
spectroscopy of \nsl\ massive stars 
located at spectroscopic distances between 1.3 and 10 kpc in the region $-30^\circ\lesssim l \lesssim+30^\circ$ and $-15^\circ\lesssim b \lesssim+15^\circ$. 
These sight lines probe gas above and below the Sagittarius Carina, Scutum Crux-Centaurus, Norma, and Near 3 kpc spiral arms in a range of $z$-height 
from 0 to 1.5 kpc. 
Along the \nsl\ sight lines, we measure velocity centroids for \ncomp\ UV absorption-line components across multiple gas phases (molecular CO, neutral, low ion, and high ion).
We find that 619/800 components have velocities that are consistent with a simple model of co-rotation with the disk, indicating that multiphase gas with disk-like 
kinematics extends at least 1 kpc into the halo.
We present a database of absorption-line parameters that
can be used for kinematic modeling of gas flows into and out of the Galactic disk.
\end{abstract}


\section{Introduction} \label{sec:intro}

\begin{figure*}[t!]
    \centering
    \includegraphics[width=\textwidth]{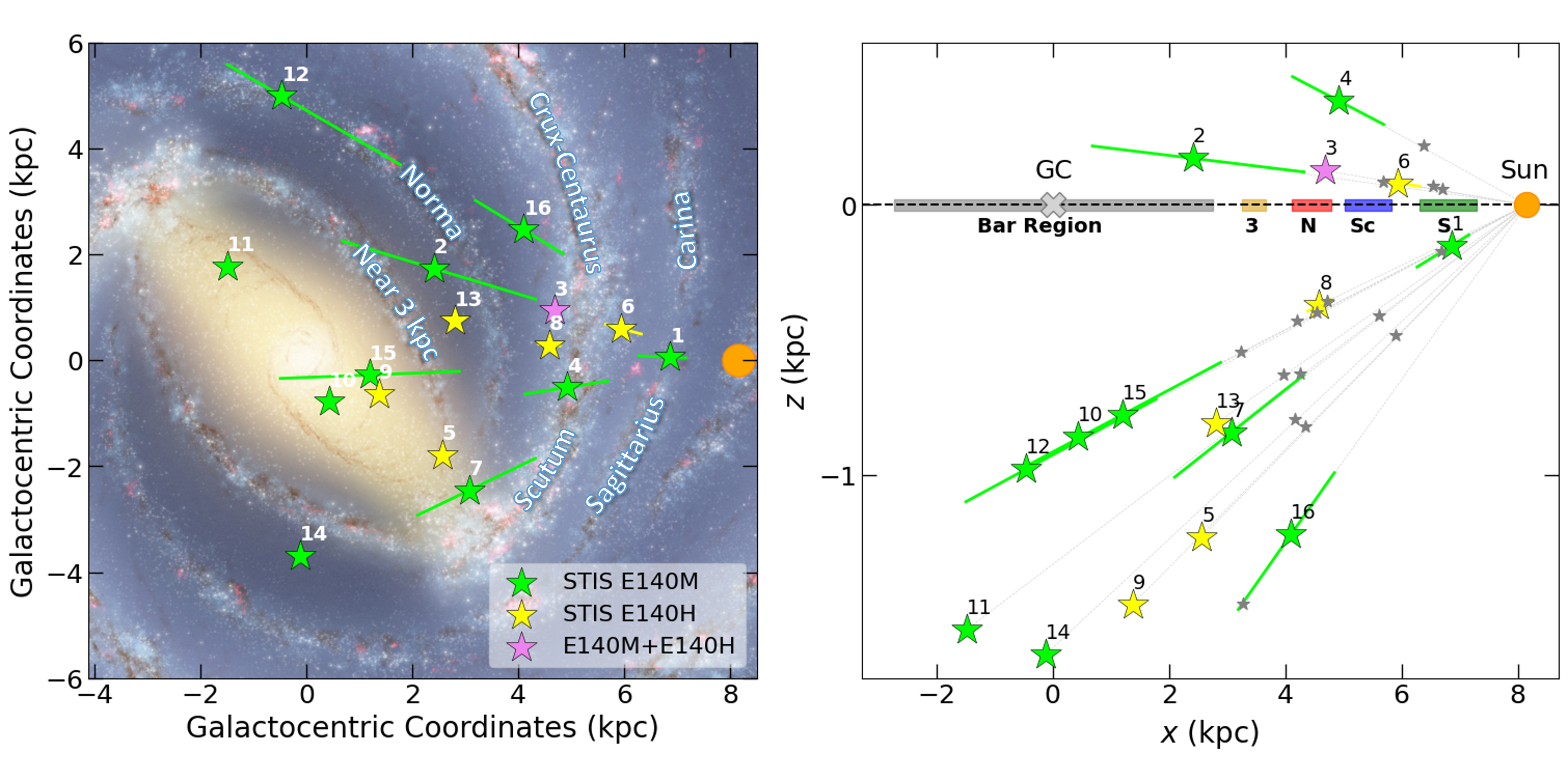}
    \caption{Left panel: Top-down artist’s conception of the Inner Milky Way showing the four main spiral arms. The location of the Sun is marked with a orange circle and each archival HST stellar target is marked by a green, yellow, or pink star at its spectroscopic distance (green: STIS E140M, yellow: E140H, pink: E140M+E140H). These Inner-Galaxy sight lines probe above and below over a range of the 4 inner spiral arms: Sagittarius Carina (S), Scutum Crux-Centaurus (Sc), Norma (N), and Near-3-kpc (3). The number next to each star indicates the target’s identifier in Table \ref{tab:t1}. Adapted from NASA/JPL-Caltech/R. Hurt (SSC/Caltech). 
    Right panel: 
    Vertical distribution of the sample along longitude $l=0\degree$, 
    with $z$ and $x$ in Galactocentric coordinates.
    The light gray stars show the Gaia DR3 positions of the stars \citep{bailer2021}.
    }
    \label{fig:sample}
\end{figure*}

The Inner Galaxy is one of the richest parts of the sky for ISM analysis.
High column densities of molecular, neutral, and ionized gas result in a wide range of absorption lines being detectable in the ultraviolet (UV) spectra of background stars.
Furthermore, all four of the Galaxy's principal spiral arms and the Galactic bar pass between the Sun and the GC, so that sight lines toward Inner Galaxy targets are complex and probe gas flows above and below many foreground structures.

Many prior UV absorption-line analyses of interstellar gas in the Milky Way have focused on individual sight lines 
\citep[e.g.,][]{savage1990, savage1994, savage2001, sembach1991,sembach1993,sembach1994, sterling2002,knauth2003,fox2003}. 
Surveys of \ion{O}{6} \citep{bowen2008} and other high ions 
\citep{lehner2011} also exist, but no surveys have specifically targeted the Galactic Center region.
In this paper we present such a targeted study, focusing on the UV properties of the ISM in, above, and below the midplane of the Inner Galaxy.
Of particular interest are the kinematics and vertical extent of the multiphase gas, which can be used to trace gas flows into and out of the Galactic disk.

The paper is organized as follows. 
In Section \ref{sec:meth} we describe our sample selection and
data handling procedures.
In Section \ref{sec:hd177989} we present the UV spectra for one sight line, to illustrate the data quality and analysis methods.
We present our results and a discussion in Section \ref{sec:results}. 
In Section \ref{sec:sum} we summarize our results.

\begin{deluxetable*}{llccccccl}[t!]
\tablecaption{The Inner-Galaxy Sample of OB stars}
\tablehead{
\colhead{ID} & \colhead{Star} & \colhead{$l$ (\degree)} & \colhead{$b$ (\degree)} & \colhead{$d_\mathrm{Gaia}$ (kpc)$^a$} & \colhead{$d_\mathrm{spec}$ (kpc)} & \colhead{Ref.$^b$} & \colhead{Region$^c$} & \colhead{Archival Data Sets$^d$; Program ID}
}
\startdata 
1 & HD 165955 & 357.41 & $-$7.43 & 1.47$^{+0.12}_{-0.09}$ & 1.3$^{+0.6}_{-0.3}$ & 1 & S & E140M (1); 8662 \\
2 & HD 151805 & 343.20 & 1.59 & 1.51$^{+0.07}_{-0.06}$ & 6.0$^{+1.8}_{-2.0}$ & 1 & S Sc N 3 & E140M (1); 9434 \\
3 & HD 152590 & 344.84 & 1.83 & 1.68$^{+0.08}_{-0.06}$ & 3.60 & 2 & S Sc N & E140M (1); E140H (3@1271) \\
  &           &        &      &  &&&& E230H (4@2263); 9434, 9855, 8241, 9465 \\
4 & HD 160641 & 8.99 & 6.49 & 1.79$^{+0.09}_{-0.10}$ & 3.3$\pm$0.8 & 3 & S Sc N 3 & E140M (1); 8603 \\
5 & HD 177989 & 17.81 & $-$11.88 & 2.41$^{+0.20}_{-0.19}$ & 6.0 & 2 & S Sc N & E140H (2@1271;1@1453;2@1489) \\
  &           &        &      &  &&&& E230H (5@2713); 7270, 7087 \\
6 & HD 152723 & 344.81 & 1.61 & 2.55$^{+0.80}_{-0.87}$ & 2.3$^{+0.3}_{-0.4}$ & 1 & S Sc & E140H (1@1307;1@1489) \\
  &           &        &      &  &&&& E230H (1@2263); 8662, 16285 \\
7 & HD 178487 & 25.78 & $-$8.56 & 2.85$^{+0.36}_{-0.25}$ & 5.7$^{+1.5}_{-1.4}$ & 1 & S Sc N & E140M (1); 13448 \\
8 & HD 163758 & 355.36 & $-$6.10 & 3.46$^{+0.38}_{-0.47}$ & 3.60$\pm$0.20 & 4 & S Sc N & E140H (1@1489); 8662 \\
9 & HD 173502 & 5.36 & $-$12.27 & 3.91$^{+0.63}_{-0.58}$ & 6.97 & 5 & S Sc N 3 & E140H (1@1271); 12192 \\
10 & HD 168941 & 5.82 & $-$6.31 & 4.00$^{+0.60}_{-0.53}$ & 7.80 & 2 & S Sc N 3 & E140M (1); 9434 \\
11 & HD 163522 & 349.57 & $-$9.09 & 4.01$^{+0.56}_{-0.47}$ & 9.92 & 2 & S Sc N 3 & E140M (1); 13448 \\
12 & HD 148422 & 329.92 & $-$5.60 & 4.18$^{+0.52}_{-0.31}$ & 10.0$^{+1.2}_{-2.6}$ & 1 & S Sc N & E140M (1); 13448 \\
13 & HD 164340 & 352.06 & $-$8.60 & 4.26$^{+1.25}_{-0.63}$ & 5.46 & 5 & S Sc N 3 & E140H (1@1271); 8241 \\
14 & HD 179407 & 24.02 & $-$10.4 & 4.44$^{+0.69}_{-0.47}$ & 9.21 & 2 & S Sc N & E140M (1); 13448 \\
15 & HD 167402 & 2.26 & $-$6.39 & 4.94$^{+0.83}_{-0.73}$ & 7.0$\pm$1.7 & 6 & S Sc N 3 & E140M (1); 13448 \\
16 & HD 156359 & 328.68 & $-$14.52 & 5.91$^{+0.16}_{-0.23}$ & 4.9$^{+1.1}_{-0.9}$ & 7 & S Sc N & E140M (1); 9434 \\
\enddata
\tablecomments{ \\
$^a$ Gaia DR3 parallax-based distances from \citet{bailer2021}. \\
$^b$ References for spectroscopic distances: (1) \citet{bowen2008}; (2) \citet{jenkins2009} no distance range provided; (3) \citet{lynas1987}; (4) \citet{bouret2012}; (5) \citet{sembach1993} no distance range provided; (6) \citet{savage2017}; (7) priv. comm. J. Ma\'{i}z-Apell\'{a}niz (8) \citet{ryans1997} \\
$^c$ Region above or below the Galactic plane probed by
the upper bound of the spectroscopic distance. Designation is S=Sagittarius Carina, Sc=Scutum Crux-Centaurus, N=Norma, 3=Near 3 kpc. \\
$^d$ HST--STIS FUV datasets. The central wavelength of all E140M archival spectra is $\lambda$1425. Numbers in parentheses are numbers of exposures at the central wavelength in Angstroms. \\
}
\label{tab:t1}
\end{deluxetable*}  

\section{Data and Methods} \label{sec:meth}
\subsection{Sample Selection}
We formed a sample of Inner-Galaxy sight lines by selecting STIS massive-star spectra from the MAST archive in a sky region surrounding the Galactic Center, defined by $330\degree\lesssim l \lesssim30\degree$ and $-15\degree\lesssim b \lesssim15\degree$.
This selection identified \nsl\ O and B-type stars in the Inner Galaxy region with STIS FUV spectra using the E140M, E140H, E230M, and E230H gratings. 
This sample covers both Quadrants I ($0\degree<l<90\degree$) and IV ($270\degree<l<360\degree$) of the Galaxy, with an even distribution of 
sight lines across these two quadrants (see Figure~\ref{fig:sample}).

We use two sets of distance estimates for our target stars: Gaia DR3 parallax-based distances \citep[][]{bailer2021} and spectroscopic distances from the literature. 
These distances often differ substantially, so we consider both distances when interpreting foreground gas components in the stellar spectra. 
Table \ref{tab:t1} provides the coordinate and distance information for each star, as well as the grating and setting for each dataset.

\subsection{STIS Ultraviolet Spectra}
We downloaded archival STIS UV spectra of the \nsl\ sight lines from the MAST HST archive and reduced them using the standard STIS data-reduction pipeline \texttt{calstis} (v.3.4.2, \citealt{dressel2007}). 
We combined the echelle orders to create a single continuous spectrum. In regions of order-overlap spectral counts were combined to increase the signal-to-noise ratio (SNR).
The data have SNR $\sim$10--38 per resolution element and a FWHM velocity resolution of 2.5 and 6.5 km s$^{-1}$ for E140H/E230H and E140M/E230M, respectively. 

We identified interstellar absorption lines in the spectra and fit continua locally around each line using the Python package \texttt{linetools} \citep{prochaska2017}. 
The identified absorption lines were measured using the Voigt profile fitting software, \textsc{VPFIT} (v.12.2; \citealt{carswell2014}) with initial estimates for fit parameters (e.g., line centroid, column density, Doppler parameter, and number of components) made using \textsc{RDGEN}, which is included with \textsc{VPFIT}. 
Both programs use wavelengths and oscillator strengths from the compilations of \citet{morton2003} and \citet{cashman2017}.

During the fitting process, we employed the HST/STIS line spread functions for the various gratings used to account for instrumental line broadening. 
All absorption components were fit independently. 
In cases where transitions showed saturation or severe blending, the velocity centroid and Doppler $b$-values of the affected components were tied to those from an unsaturated transition of the same ionization stage. The saturated components are presented as lower limits. See Section \ref{sec:fitting} in the Appendix for a full description of the Voigt profile fitting process that we have adopted for our analysis.
All wavelengths and velocities for the absorption-line features in this paper are given in the LSR reference frame.

\subsection{Model of Differential Co-rotation} \label{subsec:corotation}
In each sight line, we calculate the range of velocities that can be explained by a model of differential Galactic rotation \citep[see][]{wakker1991}; similar models have been used in numerous CGM studies to distinguish clouds with high peculiar velocities or high-velocity clouds from disk gas \citep[e.g.,][]{sembach2003, wakker2008, richter2017, cashman2021}. 
The model uses a flat rotation curve with $v$=220\kms\ from $R$=0.5 to 26 kpc, 
with the Sun at $R$=8.0 kpc. This disk has a thickness of 2 kpc at $R$=1 kpc, 
increasing linearly to 6 kpc at $R$=26 kpc. 
Inside 0.5 kpc the disk is 2 kpc thick and has solid-body rotation,
so that $v(R) = 220 (R/0.5\,{\rm kpc})$\kms.
For a given latitude and longitude, the highest and lowest velocity given by the model provides a range of allowed velocities for a cloud velocity to be co-rotating with the disk.

\section{The HD 177989 sight line}\label{sec:hd177989}

\begin{figure}[!t]
    \centering
    \includegraphics[width=\columnwidth]{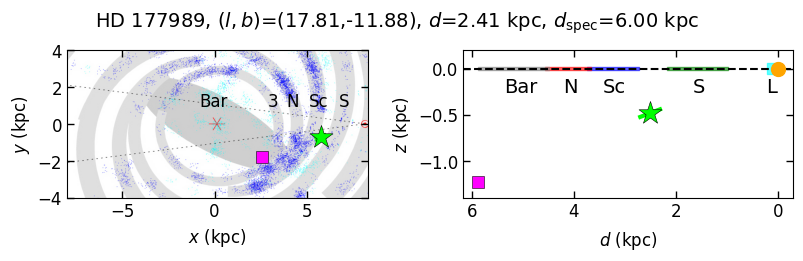}
    \includegraphics[width=\columnwidth]{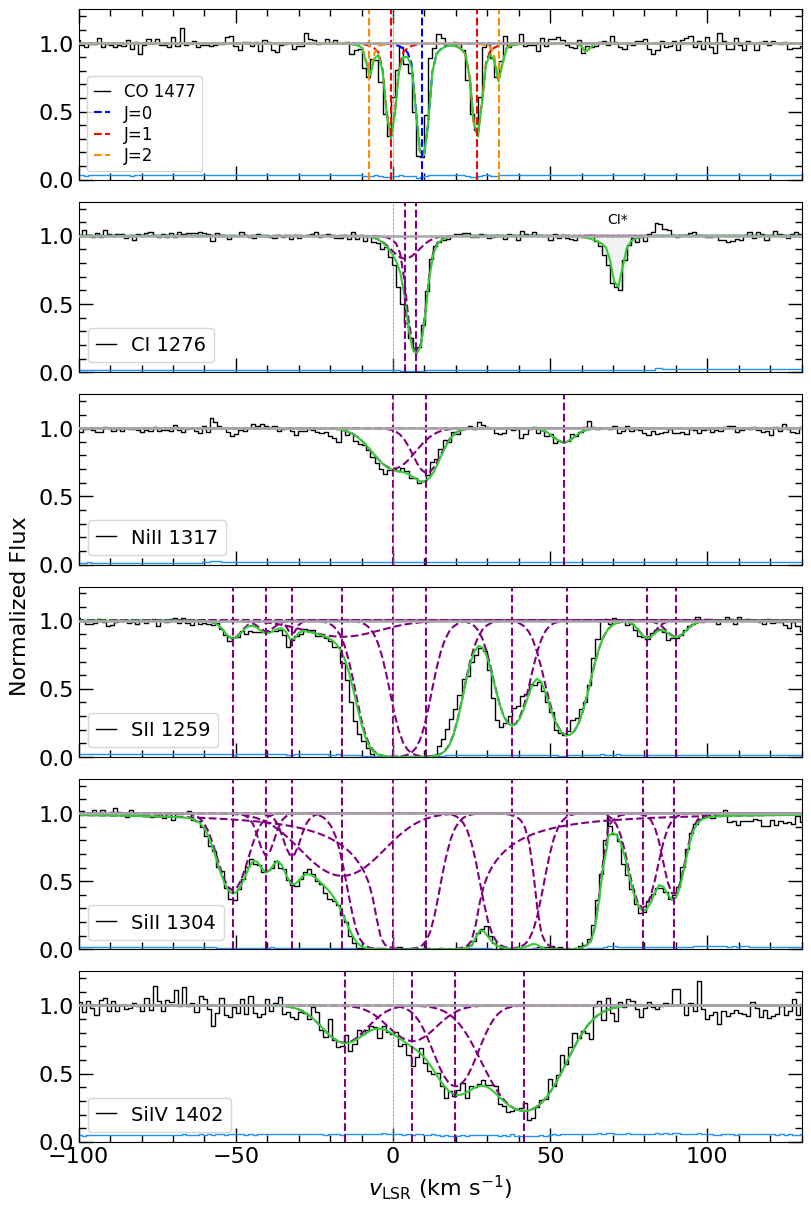}
    \caption{UV absorption-line analysis of the HST/STIS E140H spectrum for HD 177989. Top-left: overhead view from \citetalias{reid2019} overlaid with the star's Gaia distance (green star) and spectroscopic position (magenta square) from \citet{jenkins2009}. Top-right: side view of the star's position relative to the spiral arm midplane (horizontal bars) as traced by the maser data from \citetalias{reid2019}: Sagittarius (S--green), Scutum (Sc--blue), and Norma (N--red). The inward extent of the Local Spur (cyan) from the position of the Sun (orange circle) and the Bar region (gray) are shown. Bottom: continuum-normalized flux (black) versus \vlsr\ for six UV metal-line transitions. 
    Individual and overall Voigt profile fits to the data are the dashed purple and green curves, respectively, with component centroids marked with dashed purple vertical lines.}
    \label{fig:hd177989}
\end{figure}

To illustrate our technique of using UV absorption-line spectroscopy to probe the inner Galaxy region, in this section we present a detailed account of the multiphase gas toward target star HD\,177989.
At a latitude of $b=-11.88^{\circ}$ and spectroscopic distance of $d=6.00$ kpc \citep{jenkins2009}, this star is distant enough to probe beneath multiple spiral arms in Quadrant I. 
At this longitude, differential Galactic rotation, as discussed in Section \ref{subsec:corotation}, gives a range of $0 < v$(\kms) $<149$ for co-rotation with the disk.
We provide detailed descriptions of the multiphase gas seen toward all other sample sight lines in Appendix~\ref{app:indiv}.

Figure \ref{fig:hd177989} shows a stack plot for the HD~177989 sight line together with two panels on top that illustrate the location of the star relative to the Galactic disk.
The top-left panel shows a top-down view of the Galaxy, with the spiral-arm structure taken from \citetalias{reid2019}. 
This arm structure is observationally based on the locations of high-mass star-forming regions with measured trigonometric parallaxes (dark blue points). 
The width of the arms are denoted by the gray paths, which \citetalias{reid2019} defined to enclose 90\% of their sources. 
There are two locations shown for the background target star, one based on the Gaia DR3 parallax distance (green star, \citealt{bailer2021}) and one based on a spectroscopic distance (pink square, \citealt{jenkins2009}).
The top-right panel shows a $z$-height side view where the $X$ and $Y$ coordinates are converted into a radial distance $d=(X^2+Y^2)^{1/2}$.
The colored horizontal bars show the midplane spiral arm widths in the thin disk along the line of sight to the Galactic Center, where the height corresponds to the average $z$ height of the maser emission determined by \citetalias{reid2019} of $\sim$ 20 pc.
The six lower panels in Figure \ref{fig:hd177989} show a variety of multiphase UV metal lines detected in the HD\,177989 spectrum. All spectra are continuum-normalized.  

HD\,177989 is a well-studied sight line passing through the high-latitude ejecta of the Scutum supershell \citep[GS 018-04+44;][]{sterling2002}. 
The STIS E140H spectrum has exceptionally high signal-to-noise, which is especially evident in the CO 1477 {\AA} region shown in Figure \ref{fig:hd177989}, which shows resolved $J=0,1,2$ lines. Given that molecular CO is largely confined to the thin disk, we conclude that the observed CO at \vlsr\ = 9.6 \kms\ arises from our Local Orion Spur, since \citetalias{reid2019} observe molecular emission associated with Sagittarius at \vlsr\ = 19.4 \kms. Furthermore, the sight line probes approximately 300 pc below Sagittarius and 680 pc below Scutum, which would place the CO well outside the thin disk.

\citet{savage2001} presented observations of very strong \ion{Si}{4} and \ion{C}{4} components that pass under the Scutum arm. 
They state that the low-and high-ionization absorption line components extending from approximately $+$30 to $+$70 \kms\ trace gas associated with the Scutum supershell outflow $\sim$700 pc below the Galactic plane. 
We also see strong absorption at this velocity range and $z$-height and note that their spiral-arm association is only possible by adopting a distance from our position that is higher than the Gaia DR3 distance of $d=2.41\pm0.20$ kpc. 
However, the spectroscopic distance calculated by \citet{jenkins2009} of $d=6.0$ kpc is distant enough to probe outflows from the disk in the region under Scutum, as well as Norma. 

\section{Results and Discussion} \label{sec:results}

\subsection{Absorption-Line Database} \label{subsec:database}

\begin{figure*}[!t]
    \centering
    \includegraphics{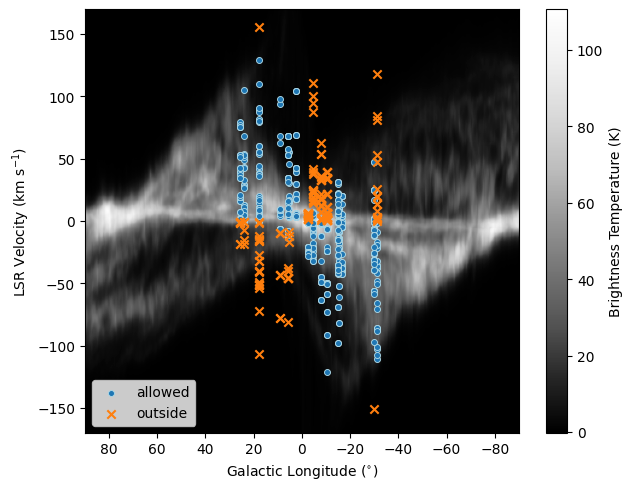}
    \caption{Location of each observed UV absorption component in velocity--longitude space. The blue circles show the components consistent with a model of differential Galactic rotation, outlined in Section \ref{sec:results}, and the orange $\times$ markers denote components outside the allowed range of rotation. The grayscale shows the \ion{H}{1} 21\,cm emission data from the LAB Survey \citep{kalberla2005} integrated over the central 10$^{\circ}$ of Galactic latitude.
    }
    \label{fig:all_data_vlsr}
\end{figure*}

Overall, we detect 800 UV absorption-line components toward the 16 inner Galaxy sight lines, across 6 different ionization states. The components are comprised of 25 molecular CO components, 233 neutral atom components (ionization state I), 419 low-ion and intermediate ions components (states II and III), and 123 high-ion components (states IV and V). The absorption line parameters are presented in Table~\ref{tab:vpfit_samp}.

We find that 619 of the 800 components (77\%) can be explained by the model of differential Galactic rotation described in Section~\ref{subsec:corotation}, i.e., the velocity centroid of the observed component falls in the range of allowed co-rotation velocities. 
Of the 619 corotating components, 20 are molecular CO, 176 are in the neutral phase, 330 are in a low- or intermediate-ion phase, and 93 are in a high-ion phase. These fractions yield corotating percentages of 80, 76, 79, and 76\%, respectively, for the four gas phases. These high fractions indicate that corotating UV gas extends to 1 kpc from the plane
for some sight lines in our sample.
In the 181 components that fall outside the range of allowed velocities, 5 are molecular CO, 57 are in the neutral phase, 89 are in a low- or intermediate-ion phase, and 30 are in a high-ion phase.

Figure \ref{fig:all_data_vlsr} presents a longitude-velocity diagram showing the 800 observed UV components coded by whether they follow the simple model for co-rotation described in Section \ref{subsec:corotation}. The data 
are overplotted on the \ion{H}{1} 21 cm emission from the LAB survey \citep{kalberla2005}. The 77\% of components that are corotating are concentrated on the \ion{H}{1} emission lobes associated with disk velocities.

\subsection{Spatial Dependence of Absorption} \label{subsec:spatial}

Our sample consists of seven stars in Galactic Quadrant I 
(Q\textsc{I}; $0^{\circ} \leq l \leq+90^{\circ}$) and nine stars in Galactic Quadrant IV (Q\textsc{IV}; $270^{\circ} \leq l < 360^{\circ}$).
Four stars are at positive latitudes between $+1.6^{\circ} \lesssim b \lesssim+6.5^{\circ}$ and twelve stars at negative latitudes between $-14.5^{\circ} \lesssim b \lesssim-5.6^{\circ}$.

\begin{figure}[!t]
    \centering
    \includegraphics[width=\columnwidth]{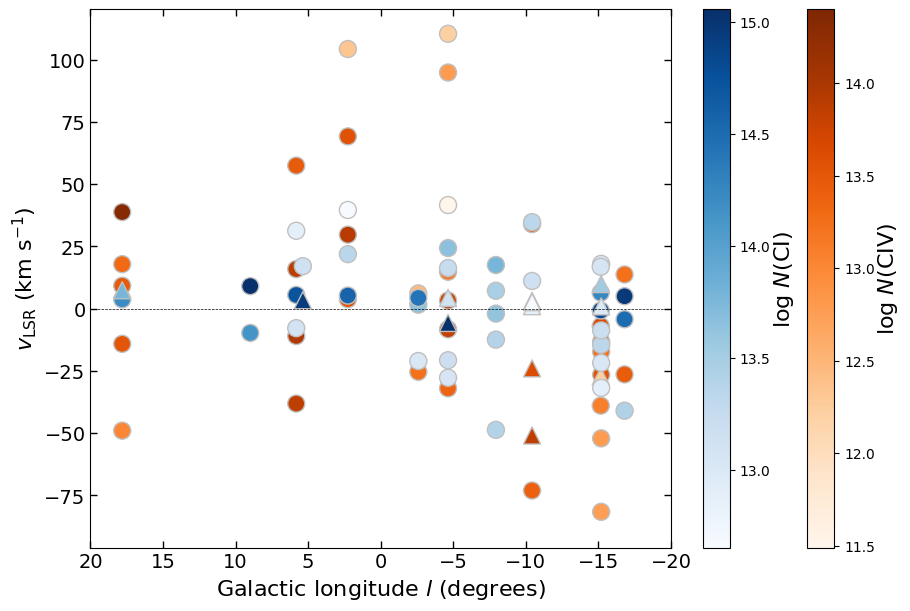}
    \includegraphics[width=\columnwidth]{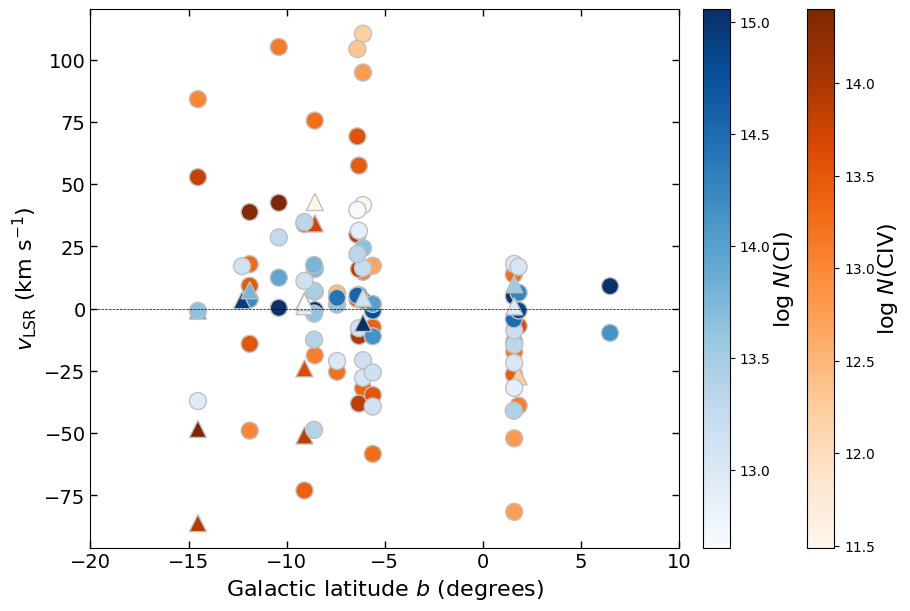}
    \caption{Top panel: LSR velocity versus Galactic longitude for observed \ion{C}{1} (blue) and \ion{C}{4} (orange) components. The gradient color bar indicates the magnitude of the column density in \cm. Each sight line contributes a series of points at the same longitude.
    Bottom panel: Same as the top panel, but for LSR velocity versus Galactic latitude. The horizontal line at $b=0\degree$ indicates the position of the midplane. Each sight line contributes a series of points at the same latitude.
    }
    \label{fig:lat-lon}
\end{figure}

We consider the dependence of UV component properties (velocity, ion stage, and column density) on longitude and latitude in Figure~\ref{fig:lat-lon}. The top panel compares LSR velocity with longitude for \ion{C}{1} and \ion{C}{4} components. The observed \ion{C}{1} components are largely confined to a velocity region of $\pm50$ \kms, with the highest column density components clustered within $\pm10$ \kms. \ion{C}{4}, however, has a broader velocity spread out to $\pm100$ \kms\ with no discernible column density dependence on velocity. We see the highest velocity \ion{C}{4} gas near $l=0\degree$ and for $l\geq10\degree$.

A comparison of LSR velocity with Galactic latitude (see lower panel of Figure \ref{fig:lat-lon}) reveals a strong trend of velocity with latitude. For the negative latitude sight lines (left half of the plot), the \ion{C}{1} components are concentrated between $b=-5$ and $-10\degree$ in a velocity window of $\pm40$ \kms, outside of which there are noticeably fewer \ion{C}{1} detections. \ion{C}{4}, on the other hand, shows a wide velocity window of $\pm100$ \kms\ between $b=-5$ and $-15\degree$ and with approximately twice as many detections as \ion{C}{1} between $b=-10$ and $-15\degree$. The number of \ion{C}{1} detections also decline as the latitude becomes more negative. Even though we only have 4 sight lines at positive latitudes, we see a similar trend in lower velocity dispersion of the \ion{C}{1} components near $b=+2.5\degree$. We have only one sight line in our sample at higher latitude near $b=+6.5\degree$ and note that we detect \ion{C}{1}, but not \ion{C}{4} or any other high ion species. We note that the STIS E140M spectrum of this sight line, HD~160641, is noisy and would benefit from future observation (see Section \ref{app:hd160641_desc} and Figure \ref{fig:hd160641}).

\begin{figure*}[!t]
    \centering
    \includegraphics[width=0.85\linewidth]{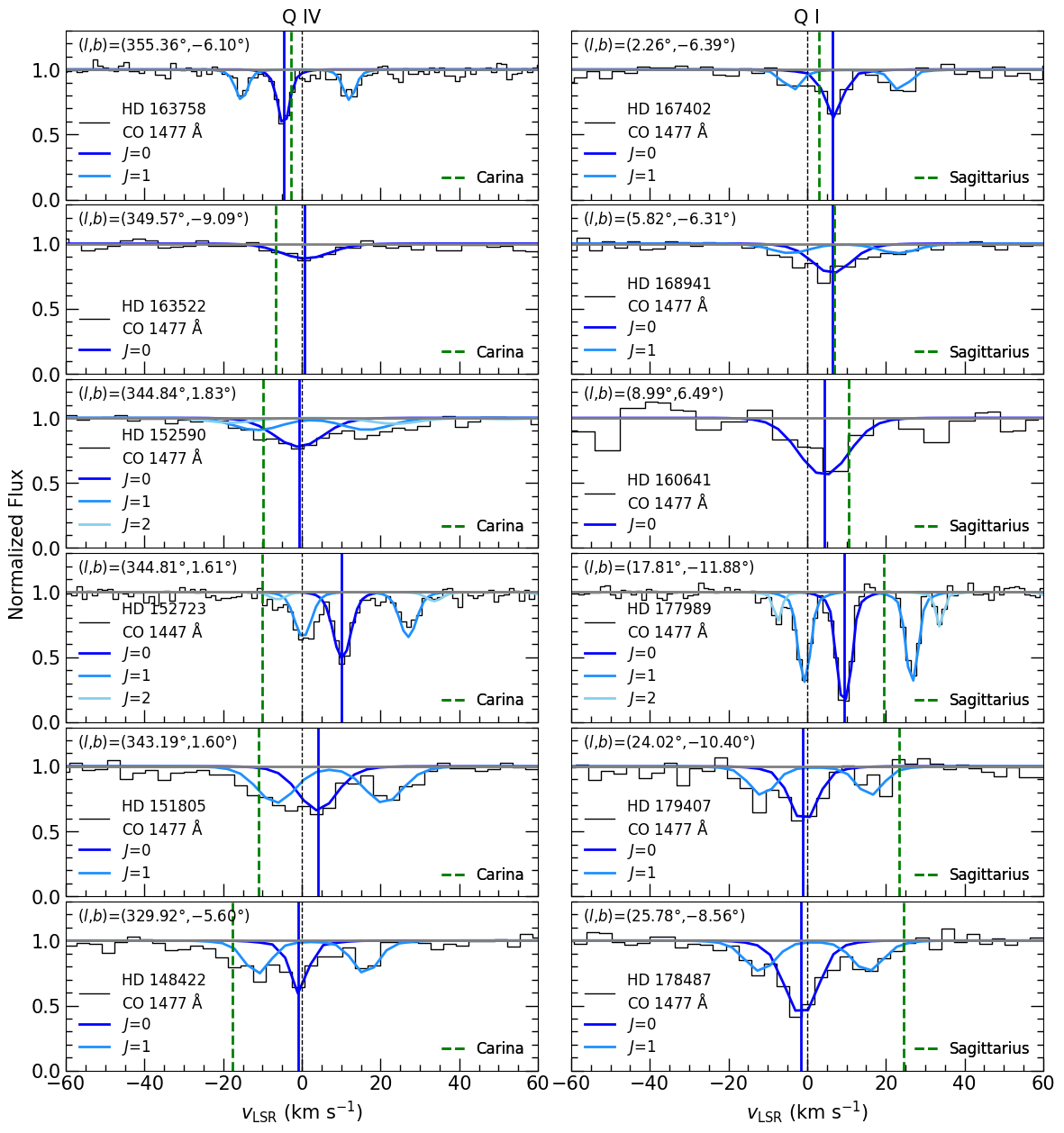}
    \caption{HST/STIS E140M or E140H spectra showing the molecular CO $\lambda$1447 or $\lambda$1477 absorption profile. CO is detected in 12 out of 16 sight lines. 
    The left panels show sight lines in Quadrant IV ($270\degree<l<360\degree$) and the
    right panels show sight lines in Quadrant I ($0\degree<l<90\degree$). The dashed green vertical line shows the peak velocity of the gas from molecular maser emission from \citetalias{reid2019} for the Sagittarius--Carina arm (left panels: Carina in Quadrant IV, right panel: Sagittarius in Quadrant I.}
    \label{fig:fig_co}
\end{figure*}

\begin{figure*}[!t]
    \centering
    \includegraphics[width=0.32\textwidth]{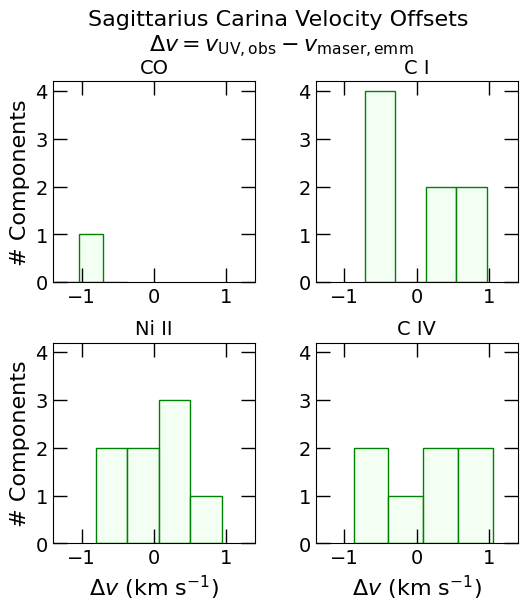}
    \includegraphics[width=0.32\textwidth]{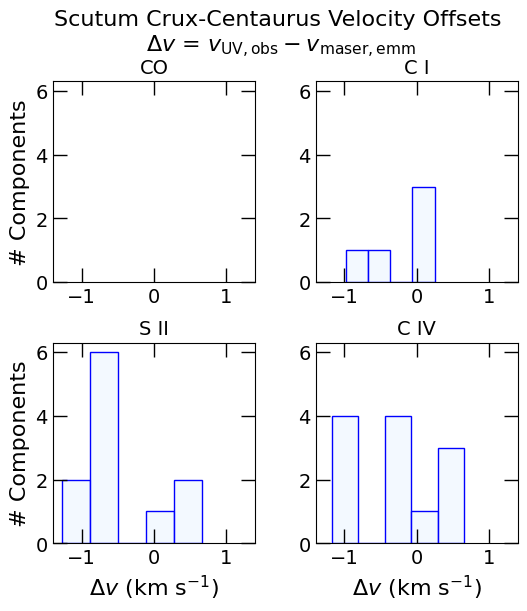}
    \includegraphics[width=0.33\textwidth]{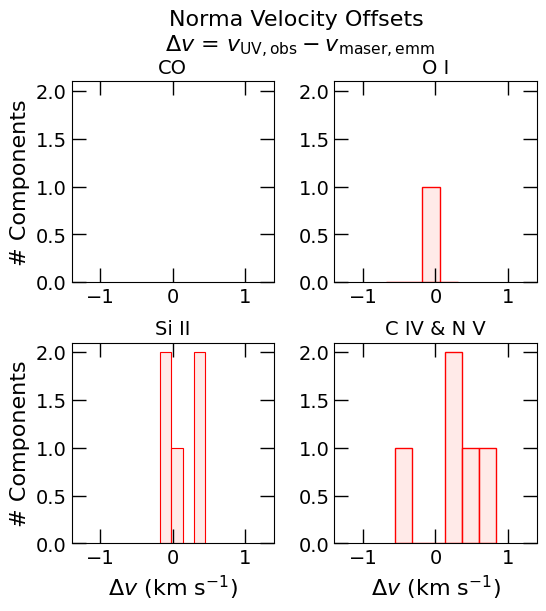}
    \caption{Distribution of UV--maser normalized velocity offsets for various gas tracers for sight lines with longitude $|l|\geq10\degree$ (see Section~\ref{subsec:spiral} for details). 
    These plots show the degree to which the UV lines from various gas phases kinematically track the spiral-arm maser emission.
    }
    \label{fig:all-vbins-hist}
\end{figure*}

\subsection{UV Molecular CO Absorption}

The high signal-to-noise of the STIS E140M and E140H UV spectra reveals molecular CO absorption in 12 out of the 16 sight lines, as shown in Figure \ref{fig:fig_co}, with 6 apiece in Q\textsc{I} and Q\textsc{IV}. 
We organized the CO data by Quadrant and then by Galactic longitude to compare the observed UV CO gas velocity centroid to the observed molecular maser emission peak velocity of the Sagittarius--Carina arm (the nearest arm to the Sun with Sagittarius in Q\textsc{I} and Carina in Q\textsc{IV}). The velocity of gas associated with the Sagittarius--Carina arm increases with increasing longitude due to Galactic rotation \citepalias[velocities from][]{reid2019}. 

The observed UV CO gas is centered near 0 \kms\ in 10/12 sight lines and is thus more likely to be associated with our local environment than with the spiral arms. The UV CO velocity centroids for sight lines HD 177989 (Q\textsc{I}) and HD 152723 (Q\textsc{IV}) are offset by approximately $+$10 \kms, but are still well offset from the expected velocities of the Sagittarius--Carina arm, suggesting that these UV CO clouds are also associated with the local environment. 
This allows us to constrain the $z$--height of the CO-bearing layer to 
$z\leq32$ pc for the low-latitude sight lines near 
$b\sim +1.6\degree$ to $z \leq 172$ pc for the higher latitude sight 
lines near $b\sim -11.9\degree$.

\subsection{Connection to Spiral Structure} \label{subsec:spiral}

Given the high percentage of components (77\%) with kinematics consistent
co-rotation with the Galactic disk (see Section~\ref{subsec:database}), it is reasonable to ask 
whether a kinematic association exists between the observed UV data and the spiral arms, which exist within the disk. We devised a simple method based on a nearest-neighbor approach (in velocity space) to compare the components we observe in UV absorption with the known velocities of the spiral arms in the midplane.

The nearest-neighbor approach is composed of three steps. 
First, for the longitude of each sight line in our sample, we extract the observed velocity of each spiral arm from the peak molecular maser emission from \citetalias{reid2019}.
Second, we establish a velocity range around each arm equivalent to half the velocity width between adjacent spiral arms. 
This process effectively assigns each UV component we observe to the spiral arm that is nearest in velocity space.
For targets that reside within the Norma arm, we use the velocity at its location as the upper end of the velocity width.
Third, we calculate the velocity offset (\delv) of each UV absorption component from the maser arm's peak velocity of the nearest spiral arm and normalize the offset by dividing by the velocity width of the spiral arm (these widths are typically of the order $\approx$20--30\kms). 
In order to reliably calculate the velocity offsets between the UV components and the maser emission from each spiral arm, we selected data from sight lines with longitude $|l|\geq 10\degree$, where Galactic rotation spreads out the spiral arms in velocity.
This longitude cut ensures that the velocity offset distribution of components is not artificially dominated by components stacked near 0 \kms, which would occur by including spiral-arm velocity widths that are narrow, e.g., as in Figure \ref{fig:hd167402} for HD~167402 at $l\sim+2.3\degree$.
This wider velocity window is at a minimum 5 and 2 times larger than the velocity resolution for the STIS E140H and E140M data, respectively.
These velocity offset distributions are shown in Figure \ref{fig:all-vbins-hist} for four gas phases: molecular, neutral, low-ions, and high-ion species, and for three spiral arms: Sagittarius--Carina, Scutum Crux--Centaurus, and Norma).
The bin sizes for the distributions are determined using the Sturges algorithm for small samples.

Although these distributions reveal the presence of UV gas components
at velocities broadly consistent with the spiral-arm velocities,
they do not show strong clustered peaks near 0 \kms, as would be expected if the UV gas components closely track the spiral arms in the midplane. Instead, the UV components are found within the entire velocity range of each spiral arm ($\Delta v=-1$ to 1).
In conclusion, the velocity offset distributions show no strong evidence for a kinematic connection between the UV components and the spiral arms. Such a connection has been reported along individual sight lines \citep{sembach1994, fox2003} but is not seen systematically in our larger sample.

\subsection{Stellar Distances}

We present both spectroscopic and Gaia parallax distance estimates for the stellar targets in our sample. 
In 10 out of \nsl\ sight lines, the spectroscopic distance is higher than the \emph{Gaia} DR3 parallax distance, by a factor of $\approx$2--3 in some cases.
The UV velocity profiles along individual sight lines cast insight on this distance discrepancy. For example, the sight line toward HD~177989 has been shown to probe through the Scutum Supershell \citep{sterling2002, savage2001}. At the Gaia DR3 distance of $d=2.41\pm0.20$ kpc \citep{bailer2021}, the sight line falls short of probing the region under the Scutum arm, but the region is permitted by the spectroscopic distance $d=6.0$ calculated by \citet{jenkins2009}. 
Other sight lines in our sample (including HD~152590, HD~178487, and HD~1640641) have similar distance discrepancies, and show kinematically complex UV spectra that may favor the spectroscopic distances over the Gaia distances. In conclusion, although trigonometric parallaxes yield the most accurate distances for nearby stars, we suggest taking spectroscopic distances into consideration for absorption-line studies of massive-star sight lines toward the GC. 

\section{Summary} \label{sec:sum}

We have conducted a UV interstellar absorption analysis of 
multiphase gas in the Inner Galaxy using HST/STIS/FUV spectroscopy of 
\nsl\ massive stars toward the Galactic Center.
We characterized the multiphase ISM in these sight lines in four gas
phases (molecular CO, neutral, low ion, and high ion).
Our study has led to the following results.

\begin{enumerate}

\item We measured a total of \ncomp\ UV absorption-line components along the \nsl\ sight lines, including 25 molecular CO components, 233 neutral atom components, 419 low-ion and intermediate ions components, and 123 high-ion components.

\item 619/800 components (77\%) have velocities that are consistent with a simple model of co-rotation with the disk. 
The co-rotation fraction for molecular components is the highest at 80\% (20/25). The co-rotation fractions for the neutral (176/233), low-ion (330/419), and high-ion (93/123) components are at similar percentages of 76, 79, and 76\% respectively.
This indicates that gas with disk-like kinematics extends to at least 1 kpc into the halo for all gas phases probed. 

\item We correlated the velocities of the observed UV absorption components with the known spiral-arm velocities in the midplane in each direction for Sagittarius--Carina, Scutum Crux--Centaurus, and Norma
using observed molecular maser data from \citetalias{reid2019}.
A connection is suggested by some individual sight lines, but in our 
overall sample, the UV gas shows no clear
association with the spiral arms. Instead, the UV lines display
complicated kinematics, with the majority following Galactic
rotation and others at highly non-circular velocities.

\end{enumerate}

The database of UV components presented in this paper will enable follow-up studies of the kinematics and ionization of gas flows in the Inner Galaxy. In future papers in this series, we plan to explore several of these properties in more detail, including the vertical extension of gas away from the midplane into the Galactic halo, and the presence of inflowing and outflowing components.

\begin{acknowledgments}
{\it Acknowledgments:} We gratefully acknowledge the significant contributions to this research program from the late Ed Jenkins, who passed away during the manuscript's preparation. 
The HST--STIS data presented in this paper were obtained from the Mikulski Archive for Space Telescopes (MAST) at the Space Telescope Science Institute. 
The specific observations analyzed can be accessed via \dataset[10.17909/31v2-6462]{http://dx.doi.org/10.17909/31v2-6462}.
We acknowledge support from the NASA Astrophysics Data Analysis Program (ADAP) under grant 80NSSC20K0435, {\it 3D Structure of the ISM toward the Galactic Center}, and from the Director's Discretionary research funds at STScI under grant D0001.82515.
We are grateful to Sapna Mishra and Ravi Sankrit for helpful feedback during the preparation of this manuscript. We thank Orlagh Creevey for valuable conversations on the Gaia DR3 distances to Milky Way OB stars.
\end{acknowledgments}

\facilities{HST--STIS}

\section{Software and third party data repository citations} \label{sec:cite}
\software{\texttt{astropy} \citep{astropy2018}, 
          \texttt{calSTIS} \citep{dressel2007},
          \texttt{linetools} \citep{prochaska2017},
          \textsc{VPFIT} \& \textsc{RDGEN} \citep{carswell2014}
          }
\clearpage

\appendix
\restartappendixnumbering

\section{Voigt Profile Fitting\label{sec:fitting}}\label{app:fitting}
\begin{deluxetable*}{lcccccccc}[t!]
\tablecaption{Voigt profile fitting results for ions detected along the line of sight toward HD 173502}
\tablehead{
\colhead{Ion} & \colhead{$v_\mathrm{LSR}$ (km s$^{-1}$)} & \colhead{$\sigma_v$ (km s$^{-1}$)} & \colhead{$b$ (km s$^{-1}$)} & \colhead{$\sigma_b$ (km s$^{-1}$)} & \colhead{log ($N$/cm$^{-2}$)}  & \colhead{$\sigma_{\mathrm{log} N}$} & \colhead{flag $N_\mathrm{low}$} & \colhead{flag $N_\mathrm{up}$}
}
\startdata
COJ0	&	...	&	...	&	...	&	...	&	13.50	&	...	&		&	y	\\
CI	&	3.39	&	0.19	&	1.48	&	0.14	&	16.54	&	0.05	&	y	&		\\
CI	&	17.05	&	0.53	&	2.87	&	1.16	&	13.14	&	0.05	&		&		\\
CI*	&	4.45	&	0.28	&	1.48	&	0.14	&	13.60	&	0.02	&		&		\\
CI*	&	14.93	&	7.12	&	2.87	&	1.16	&	12.20	&	0.26	&		&		\\
OI	&	4.35	&	0.54	&	1.48	&	0.14	&	17.73	&	0.07	&		&		\\
OI	&	15.86	&	1.10	&	2.87	&	1.16	&	17.44	&	0.10	&		&		\\
ClI	&	3.73	&	0.11	&	2.62	&	0.16	&	13.23	&	0.02	&		&		\\
MgII	&	3.31	&	0.19	&	2.93	&	0.36	&	15.80	&	0.02	&		&		\\
MgII	&	16.37	&	0.56	&	4.53	&	1.09	&	15.25	&	0.04	&		&		\\
SII	&	$-$12.44	&	0.54	&	7.93	&	0.52	&	15.13	&	0.03	&		&		\\
SII	&	9.62	&	0.35	&	4.37	&	0.14	&	19.17	&	0.14	&	y	&		\\
SII	&	30.50	&	0.74	&	5.33	&	1.23	&	14.55	&	0.08	&		&		\\
SII	&	55.14	&	0.98	&	18.86	&	1.37	&	14.84	&	0.03	&		&		\\
MnII	&	5.86	&	0.45	&	4.85	&	1.00	&	13.44	&	0.04	&		&		\\
MnII	&	20.06	&	1.18	&	6.12	&	1.29	&	12.99	&	0.07	&		&		\\
NiII	&	$-$17.20	&	1.07	&	3.72	&	2.12	&	12.66	&	0.10	&		&		\\
NiII	&	2.90	&	0.77	&	4.85	&	1.00	&	13.36	&	0.03	&		&		\\
NiII	&	15.48	&	0.99	&	6.12	&	1.29	&	13.36	&	0.03	&		&		\\
NV	&	$-$9.04	&	4.84	&	30.77	&	3.85	&	13.53	&	0.12	&		&		\\
NV	&	46.00	&	8.00	&	39.12	&	7.65	&	13.56	&	0.11	&		&		
\enddata
\tablecomments{This table is available for all sight lines in machine-readable form. The flags ``low" and ``up" indicate if log $N$ is a lower limit (saturated) or a 3$\sigma$ upper limit (non-detection).}
\label{tab:vpfit_samp}
\end{deluxetable*}

In this section we describe the Voigt profile fitting process we adopted to determine the velocity centroids and Doppler parameters ($b$--value) of the components to achieve a consistent analysis across all of our sight lines

We utilize the available instrument line spread functions for the individual HST/STIS E140M, E140H, E230M, and E230H settings provided by the Space Telescope Science Institute. Unblended and resolved components are fit independently. For transitions which show highly blended or saturated components, we first select the weakest transition in an ion stage which displays distinct structure and identify the least number of components for a best fit. We then adopt these velocity centroids for blended transitions for other species of the same ion stage if they show an overall similar profile. Transitions of the same ion stage that are too saturated or blended to discern whether or not they have similar structure to weaker ions of the same stage are excluded from fitting. 

For example, in Figure \ref{fig:hd165955}, the weakest low ion \ion{Ni}{2} 1370 {\AA} requires three components for a best fit. We adopt the velocity centroids of these three components for the much stronger \ion{S}{2} 1250 and \ion{Si}{2} 1304 transitions, and allow the $b$--value and column density $N$ to be free parameters. For saturated components, the resulting $N$ is considered a lower limit. For blended high ion transitions, we use the low ion profile structure as starting conditions if their profiles show a similar structure, and allow the parameters to vary. Otherwise, the high ion components are fit independently. 

We fit every transition possible in the spectrum according to this process, although we only show a subset of transitions in the figures. We chose the transitions in the figures to best highlight, where possible, the full range of molecular, low, intermediate, and high ions present along the line of sight. The complete results of the Voigt profile fitting are available in a machine-readable table. A portion of the table is shown in Table \ref{tab:vpfit_samp} for the HD 173502 sight line.

\section{Notes on Individual Sight Lines}\label{app:indiv}

In this section we describe the absorption features seen in each of the \nsl\ sight lines in our sample (except HD 177989, which is discussed in Section~\ref{sec:hd177989}). The sight lines are listed in the same order as Table \ref{tab:t1}, by increasing distance from the Sun according to the Gaia distances. 
The term ``Local" pertains to the Local Spur segment housing our location and the Sun.

\subsection{HD~165955}\label{app:hd165955_desc}
\begin{figure}[t!]
    \centering
    \includegraphics[width=\columnwidth]{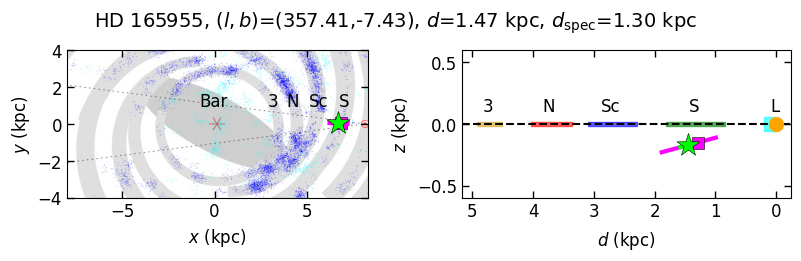}
    \includegraphics[width=\columnwidth]{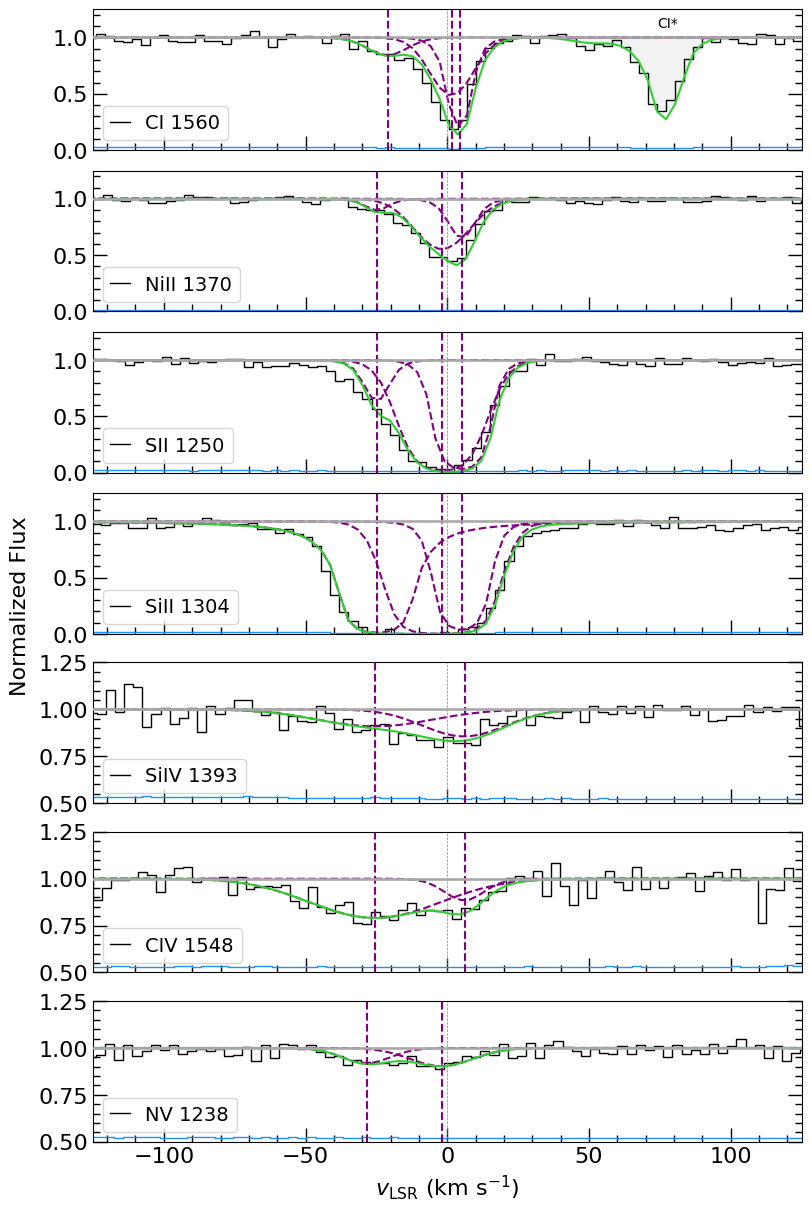}
    \caption{UV absorption-line analysis of the HST/STIS E140H UV spectrum for HD 165955 in Quadrant I. Top-left: overhead view from R19 overlaid with the star’s Gaia distance (green star) of $d=1.47^{+0.12}_{-0.09}$ kpc \citet{bailer2021} and spectroscopic position (magenta square) of $d=1.3^{+0.6}_{-0.3}$ kpc from \citet{bowen2008}. Top-right: side view of the star's position relative to the spiral arms in the Galactic plane (horizontal bars) as traced by the molecular maser data from \citetalias{reid2019}: Local Orion Spur (L--cyan), Sagittarius (S--green), Scutum (Sc--blue), and Norma (N--red). The bottom six panels show the continuum-normalized flux (black) versus \vlsr. The individual and overall Voigt profile fits to the data are shown as purple and green curves, respectively, with the component centroids marked with dashed purple lines. Regions unrelated to the transition are shaded in gray.}
    \label{fig:hd165955}
\end{figure}

\emph{HD 165955} is a high-radial-velocity star ($v=-160$ \kms, \citealt{tobin1984}) thought to have been ejected from the disk \citep{gvaramadze2009}. 
It has a Gaia DR3 distance in close agreement with its spectroscopic distance from \citet{bowen2008}, see Table \ref{tab:t1} and Figure \ref{fig:hd165955}. 
These corroborating distances indicate that the region under the Carina arm is the only arm outside our Local Spur probed along this line of sight through Quadrant IV. 
The sight line is less than 3$^\circ$ in longitude away from $l=0$, 
the lowest of our Quadrant IV sight lines. Our model of differential Galactic rotation (see Section \ref{subsec:corotation}) gives a co-rotating velocity range of $-147 \leq v$ (\kms) $ \leq 0$.
We do not observe any UV molecular CO in this sight line.  

\subsection{HD~151805}\label{app:hd151805_desc}
\begin{figure}[t!]
    \centering
    \includegraphics[width=\columnwidth]{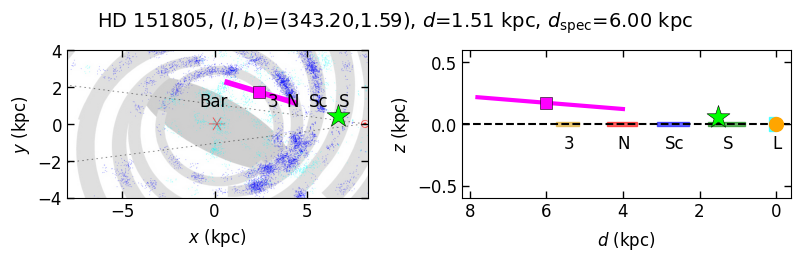}
    \includegraphics[width=\columnwidth]{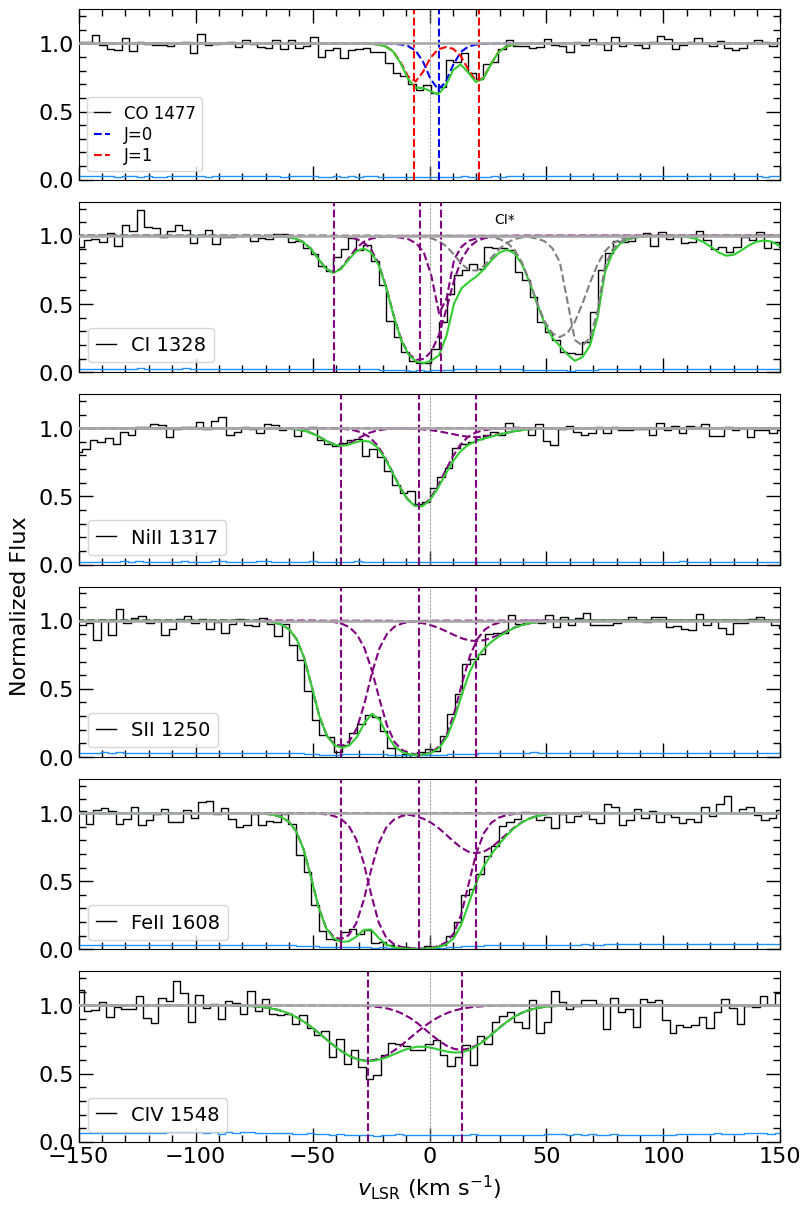}
    \caption{Similar to Figure \ref{fig:hd165955}, but for the HST--STIS E140M spectrum of HD 151805. The green star shows the Gaia DR3 distance $d=1.51^{+0.07}_{-0.06}$ kpc. The magenta square shows the spectroscopic distance of $d=6.0^{+1.8}_{-2.0}$ kpc from \citet{bowen2008}.}
    \label{fig:hd151805}
\end{figure}

\emph{HD 151805} has a Gaia DR3 distance of 1.51$^{+0.07}_{-0.06}$ kpc, well below the spectroscopic distance of 6.0$^{+1.8}_{-2.0}$ from \citet{bowen2008}. From the lower Gaia distance, only the region above the Carina arm is expected to be probed, as shown in Figure \ref{fig:hd151805}. At the higher spectroscopic distance, the region above all four arms is expected to be probed through Quadrant IV. Our model of differential Galactic rotation (see Section \ref{subsec:corotation}) gives a co-rotating velocity range of $-156 \leq v$ (\kms) $ \leq 43$.
We observe the ground-state $J=0,1$ molecular CO transitions at 1419, 1447, 1477, 1509, and 1544 {\AA} near 4 \kms. 
We note that at $l$ = 343.2$^{\circ}$ and for the higher $d_\mathrm{spec}$ = 6.00 kpc, \citetalias{reid2019} find gas near \vlsr\ = $-122.8$ \kms\ to be associated with the Near 3 kpc arm. We do not see gas in the UV at that velocity.

\subsection{HD~152590}\label{app:hd152590_desc}
\begin{figure}[t!]
    \centering
    \includegraphics[width=\columnwidth]{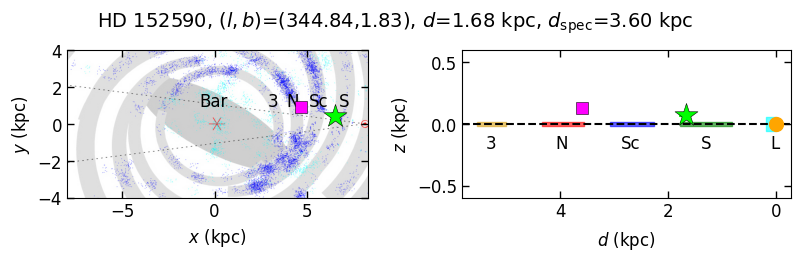}
    \includegraphics[width=\columnwidth]{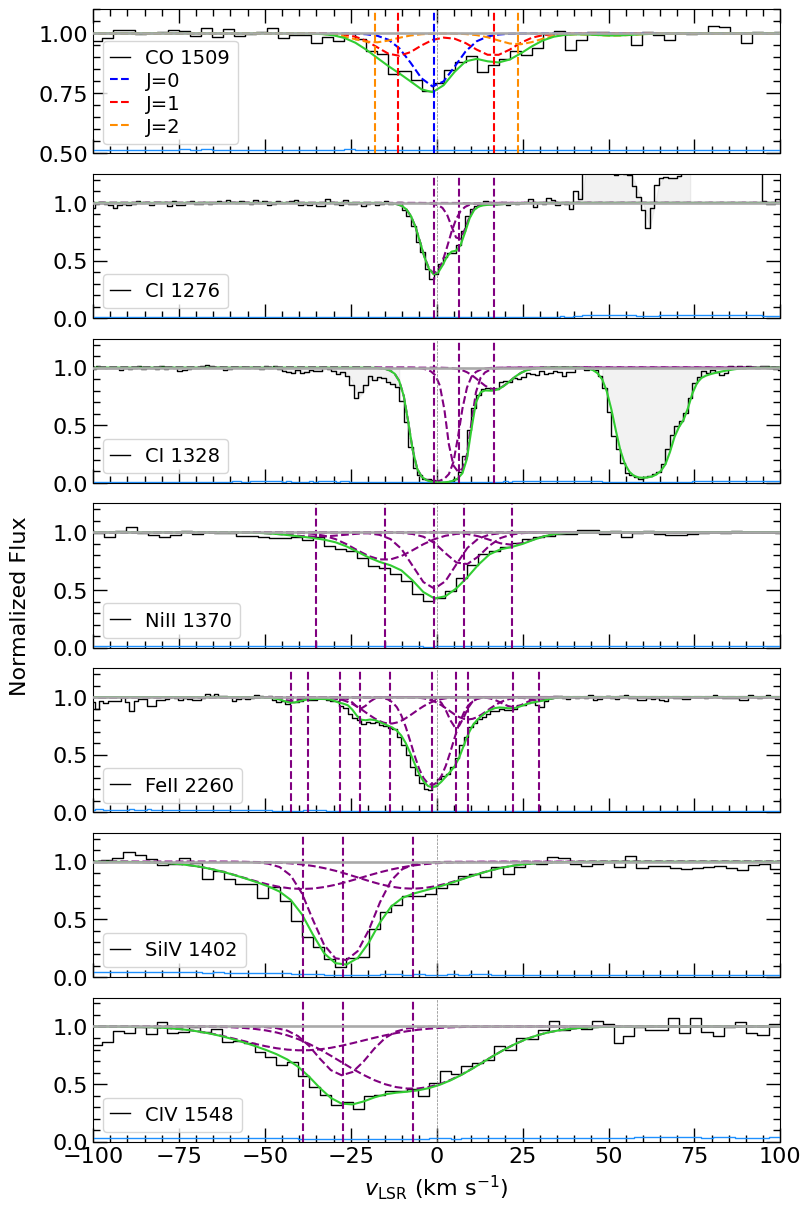}
    \caption{Same as for Figure \ref{fig:hd165955}, but for the HST--STIS E140H (\ion{C}{1}, \ion{Cl}{1}), E140M (CO, \ion{Ni}{2}, \ion{C}{4}, \ion{Si}{4}), and E230H (\ion{Fe}{2}) spectra of star HD 159590. The green star shows the Gaia DR3 distance $d=1.68^{+0.08}_{-0.06}$ kpc. The magenta square shows the spectroscopic distance of $d=3.60$ kpc from \citet{jenkins2009}.}
    \label{fig:hd152590}
\end{figure}

\emph{HD 152590} had the most available archival STIS datasets of any star in our sample, see Table \ref{tab:t1} and Figure \ref{fig:hd152590}. At the Gaia DR3 distance of 1.68$^{+0.08}_{-0.06}$ kpc, only the region above the Carina arm is expected to be probed. At the higher spectroscopic distance of 3.60 kpc from \citet{jenkins2009}  the region above both Carina and Crux-Centaurus are expected to be probed. We observe the ground-state $J=0,1$ and excited $J=2$ molecular CO transitions at 1419, 1477, 1509 {\AA} transitions near 0 \kms. Our model of differential Galactic rotation (see Section \ref{subsec:corotation}) gives a co-rotating velocity range of $-162 \leq v$ (\kms) $ \leq 39$.

\subsection{HD~160641}\label{app:hd160641_desc}
\emph{HD 160641} has an interesting history of distance study. At the Gaia DR3 distance of 1.79$^{+0.09}_{-0.10}$ kpc, only the region above the Carina arm is expected to be probed. 
Also known as V2076 Oph, HD 160641 has had its spectral type closely studied. This star was included in the $R_V$ study of Galactic O stars from the 2MASS catalog by \citet{patriarchi2003}, which classified the star's spectral type as O9.5I at a distance of 11.3 kpc. The expansive OB Catalog from \citet{reed2003} classified the star as a O9.5Iab, which resulted in a spectroscopic distance of $\sim$13.6 kpc. This catalog contains approximately 16,000 suspected OB stars, but notes that MK classification is missing for about half of them. More recently, this star has been described as an extreme Helium star (eHe) with multiple studies describing it as an eHE hot sub-dwarf \citep{lapalombara2017,schaefer2016,bobylev2015,drilling2013}. The expected distance as an eHe is $d_\mathrm{spec}$ = 3.3$\pm$0.8 kpc \citep{bobylev2015,lynas1987}. At this spectroscopic distance the region above the Carina, Crux-Centaurus, and Norma arms are expected to be probed. 

We observe the molecular CO ground-state $J=0$ transition at 1447 and 1477 {\AA}. Our model of differential Galactic rotation (see Section \ref{subsec:corotation}) gives a co-rotating velocity range of $-9 \leq v$ (\kms) $ \leq 184$.
If the star is beyond the GC (post-GC), we note that the anticipated velocities of the foreground Near 3 kpc arm from \citetalias{reid2019} is $-14.2$ \kms. 
We do observe components at this velocities in \ion{O}{1} and \ion{Si}{2}, but cannot confirm any origin. 
We present both possible distances and interpretations (nearby and post-GC) for this sight line given that the complex spectrum shows components within the velocity range for Galactic rotation. The E140M spectrum is also very noisy, so the data were binned by two pixels to increase the signal-to-noise. 

We are unable to detect any high-ion species along the line of sight, although future observations of this star at higher resolution and signal-to-noise are needed for confirmation. 

\subsection{HD~152723}\label{app:hd152723_desc}
\begin{figure}[t!]
    \centering
    \includegraphics[width=\columnwidth]{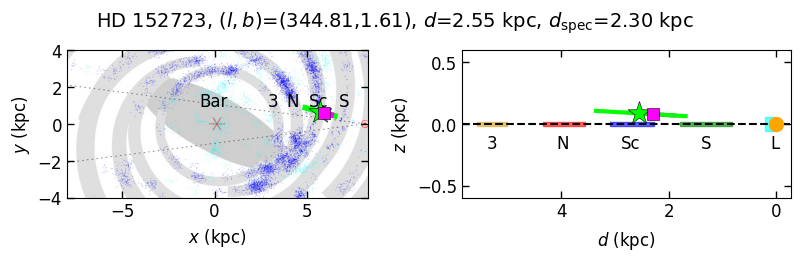}
    \includegraphics[width=\columnwidth]{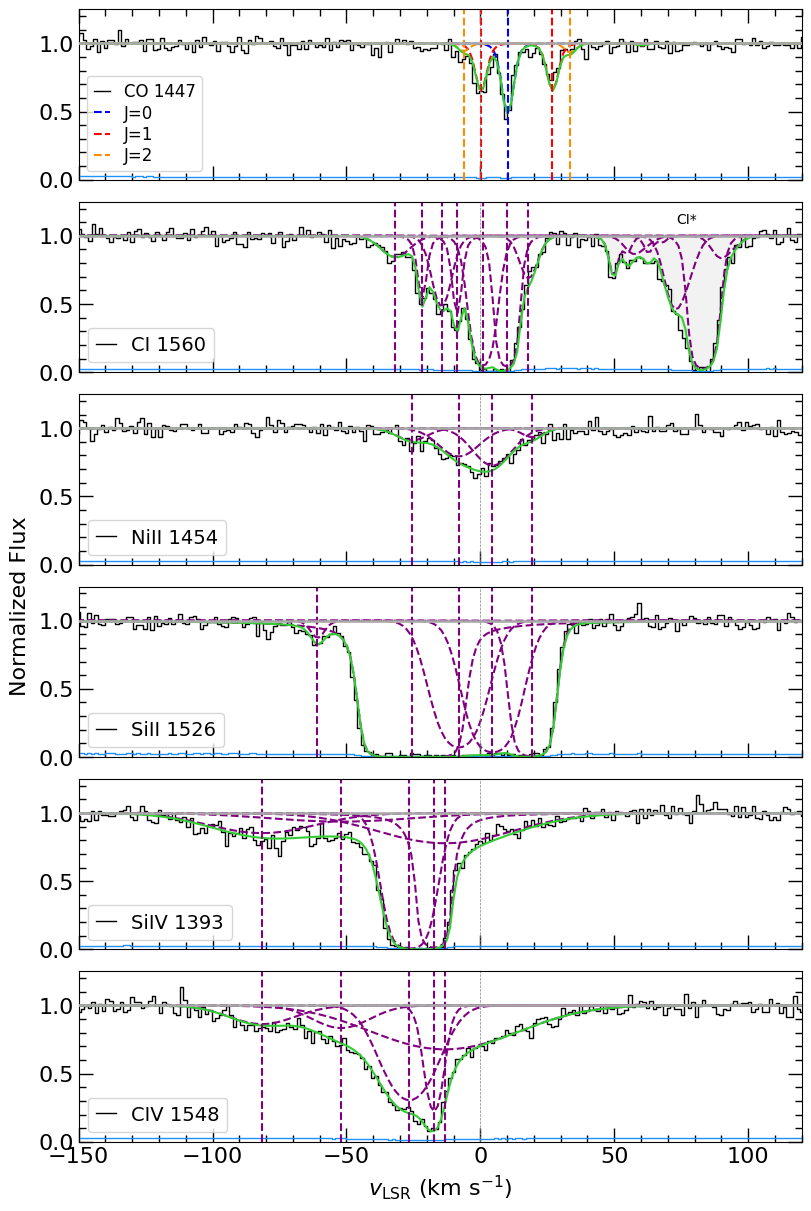}
    \caption{Same as for Figure \ref{fig:hd165955}, but for the HST--STIS E140H spectrum of HD 152723. The green star represents the Gaia DR3 distance $d=2.55^{+0.80}_{-0.87}$ kpc. The magenta square represents the spectroscopic distance $d=2.3^{+0.3}_{-0.4}$ kpc from \citet{bowen2008}.}
    \label{fig:hd152723}
\end{figure}

\emph{HD 152723} has a Gaia DR3 distance in close agreement with its spectroscopic distance from \citet{bowen2008}, as shown in Figure \ref{fig:hd152723}. This E140H spectrum is of exceptional quality as shown by the resolved CO $J=0,1,2$ 1447 transition.
Both distances place the star in or just above Crux-Centaurus in Quadrant IV.
Our model of differential Galactic rotation (see Section \ref{subsec:corotation}) gives a co-rotating velocity range of $-162 \leq v$ (\kms) $ \leq 39$.

\subsection{HD~178487}\label{app:hd178487_desc}
\begin{figure}[t!]
    \centering
    \includegraphics[width=\columnwidth]{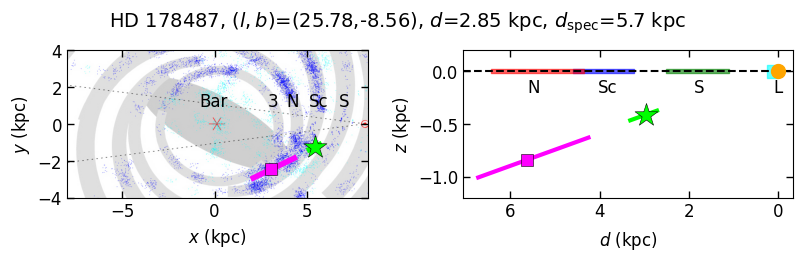}
    \includegraphics[width=\columnwidth]{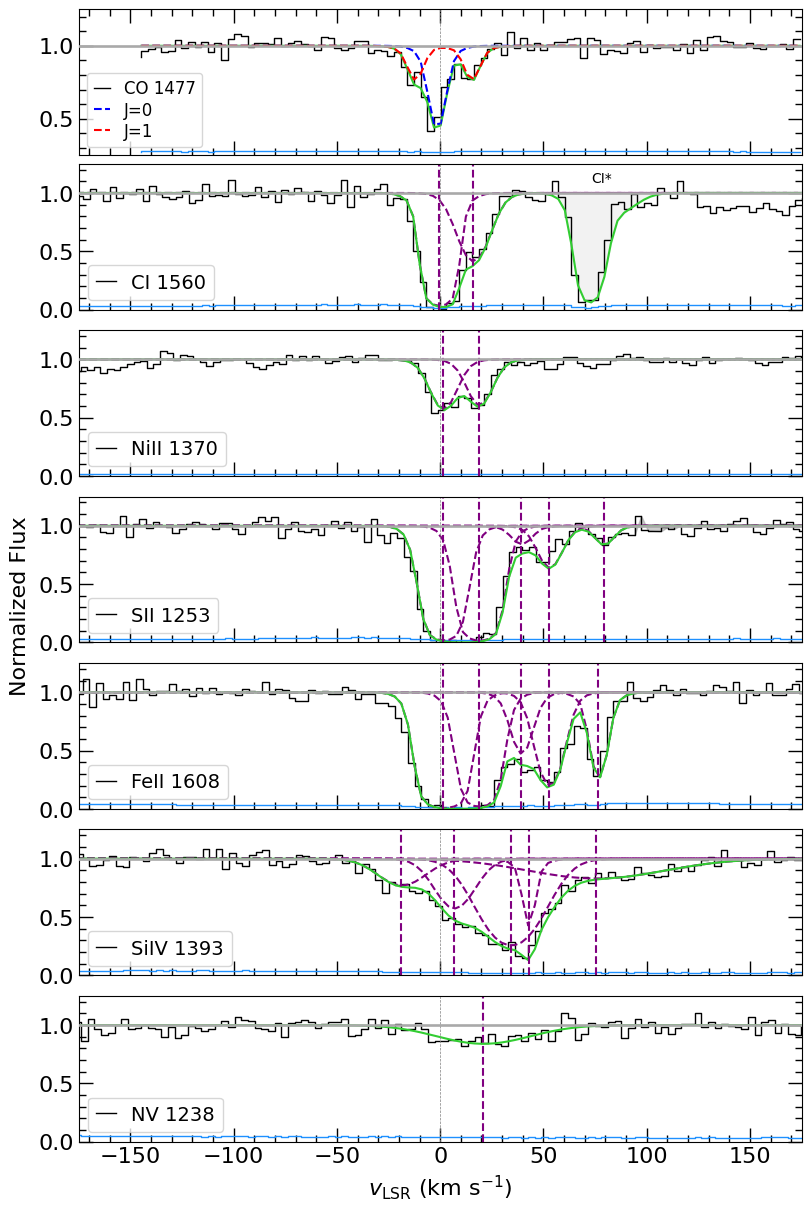}
    \caption{Same as for Figure \ref{fig:hd165955}, but for the HST--STIS E140M spectrum of HD 178487. We note that at $l=25.78^\circ$, the Near-3-kpc arm is not probed and thus its location and velocity interval cannot be shown in the top two panels. The green star represents the Gaia DR3 distance $d=2.85^{+0.36}_{-0.25}$ kpc. The magenta square represents the spectroscopic distance $d=5.7^{+1.5}_{-1.4}$ kpc from \citet{bowen2008}.}
    \label{fig:hd178487}
\end{figure}

\emph{HD 178487} has the highest longitude of the Quadrant I sight lines, see Figure \ref{fig:hd178487}. 
\citet{bowen2008} give a spectroscopic distance $d=5.7^{+1.5}_{-1.4}$ kpc, which is approximately twice the distance from the Gaia DR3 value $d=2.85^{+0.36}_{-0.25}$ kpc. At the Gaia distance, only the region under Sagittarius is probed. At the higher spectroscopic distance the region under Sagittarius, Scutum, and Norma are expected to probed.
We observe the ground-state $J=0,1$ CO $\lambda$1477 transitions. 
Our model of differential Galactic rotation (see Section \ref{subsec:corotation}) gives a co-rotating velocity range of $0 \leq v$ (\kms) $ \leq 122$.

\subsection{HD~163758}\label{app:hd163758_desc}
\begin{figure}[t!]
    \centering
    \includegraphics[width=\columnwidth]{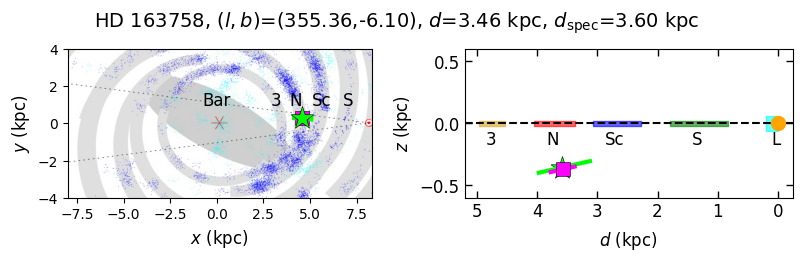}
    \includegraphics[width=\columnwidth]{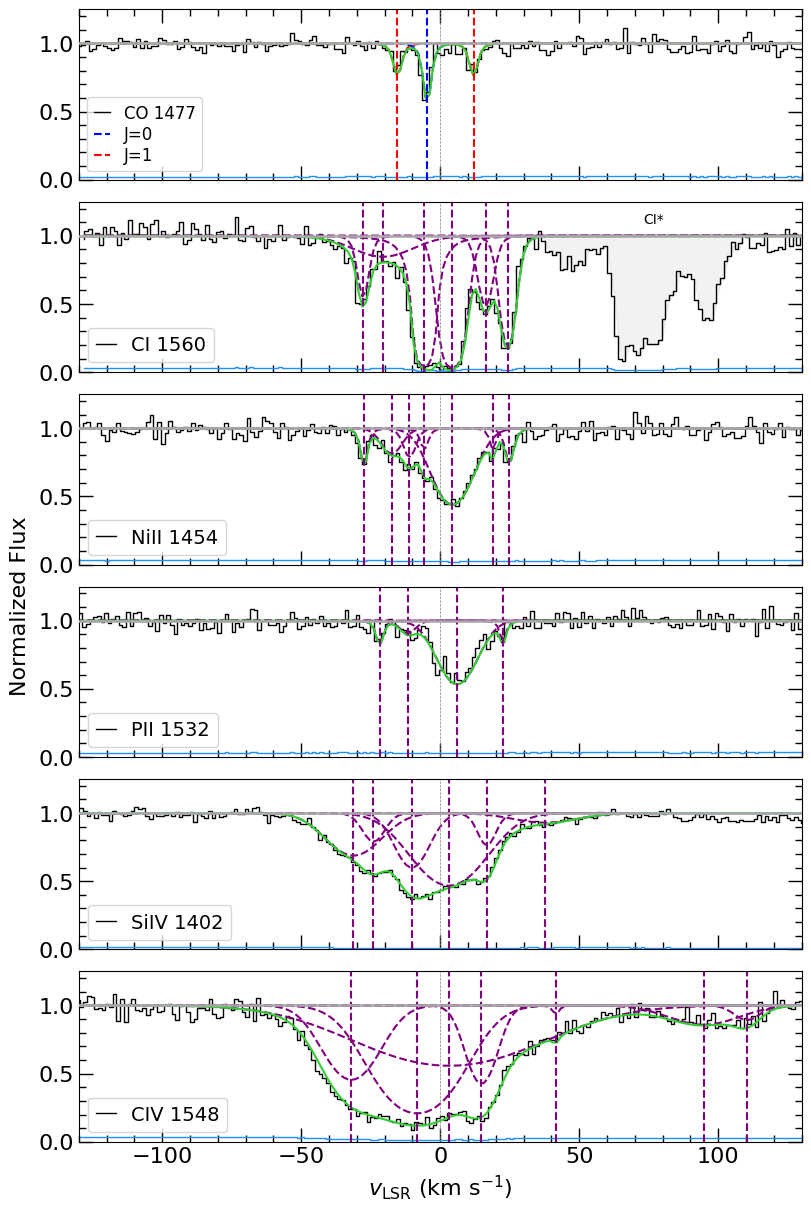}
    \caption{Same as for Figure \ref{fig:hd165955}, but for the HST--STIS E140H spectrum of HD 163758. The green star represents the Gaia DR3 distance $d=3.46^{+0.38}_{-0.47}$ kpc. The magenta square represents the spectroscopic distance $d=3.60\pm0.20$ kpc from \citet{bouret2012}.}
    \label{fig:hd163758}
\end{figure}

\emph{HD 163758} has a Gaia DR3 distance $d=3.46^{+0.38}_{-0.47}$ kpc in close agreement with its spectroscopic distance $d=3.60\pm0.20$ kpc \citep{bouret2012}, as shown in Figure \ref{fig:hd163758}. 
Both distances indicate that the regions under Carina, Crux-Centaurus, and Norma arms should be probed in Quadrant IV. 
Our model of differential Galactic rotation (see Section \ref{subsec:corotation}) gives a co-rotating velocity range of $-201 \leq v$ (\kms) $ \leq 12$.

\subsection{HD~173502}\label{app:hd173502_desc}
\emph{HD 173502} in Quadrant I has a Gaia DR3 distance $d=3.91^{+0.63}_{-0.58}$ kpc, which is appreciably shorter than it's spectroscopic distance $d=6.97$ kpc \citep{sembach1993}, but still distant enough to probe through to the region under Norma. At the higher distance, the sight line is distant enough to probe past the region under the Near 3 kpc arm.

Our model of differential Galactic rotation (see Section \ref{subsec:corotation}) gives a co-rotating velocity range of $0 \leq v$ (\kms) $ \leq 194$.
The longitude--velocity plot from \citetalias{reid2019} places the Near 3 kpc arm at a velocity of $-$26.6 \kms. We do see absorption at this velocity in low and high-ions. We cannot confirm the origin of this gas, but if associated with the Near 3 kpc feature, it would place the components approximately 1 kpc below the disk.

\subsection{HD~168941}\label{app:hd168941_desc}
\begin{figure}[t!]
    \centering
    \includegraphics[width=\columnwidth]{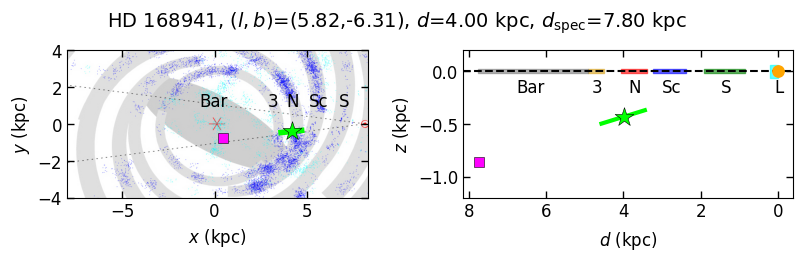}
    \includegraphics[width=\columnwidth]{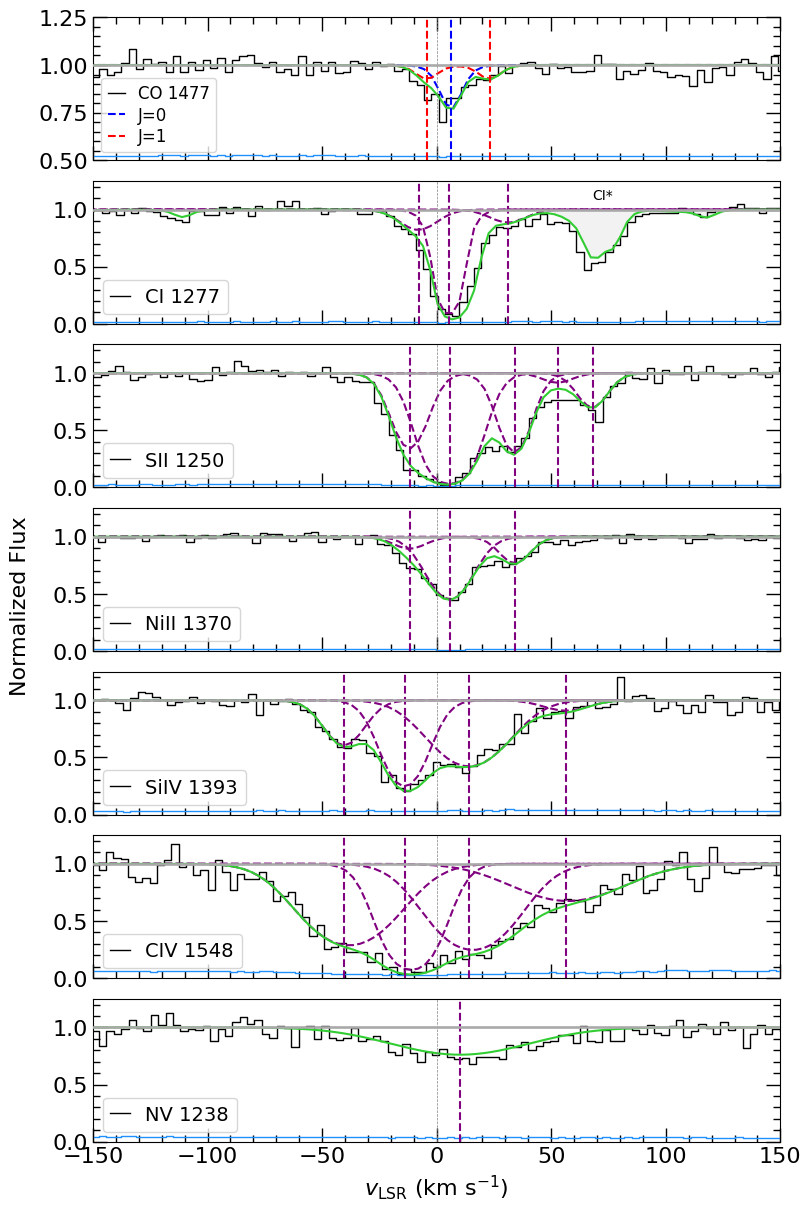}
    \caption{Same as for Figure \ref{fig:hd165955}, but for the HST--STIS E140M spectrum of HD 168941. The green star represents the Gaia DR3 distance $d=4.00^{+0.60}_{-0.53}$ kpc. The magenta square represents the spectroscopic distance $d=7.80\pm0.20$ kpc from \citet{jenkins2009}.}
    \label{fig:hd168941}
\end{figure}

\emph{HD 168941} has a Gaia DR3 distance $d=4.00^{+0.60}_{-0.53}$ that is inconsistent with a higher spectroscopic distance $d=7.80$ kpc from \citet{jenkins2009}. 
At the higher spectroscopic distance, the region under all inner spiral arms are probed as opposed to three from the upper distance end of Gaia, see Figure \ref{fig:hd168941}. We observe the ground-state $J=0,1$ CO $\lambda$1477 transitions.
Our model of differential Galactic rotation (see Section \ref{subsec:corotation}) gives a co-rotating velocity range of $-15 \leq v$ (\kms) $ \leq 196$.
For this sight line longitude, \citetalias{reid2019} places the Near 3 kpc arm at a velocity of $-$25.0 \kms. We do observe \ion{Si}{4}, \ion{C}{4}, and \ion{N}{5} at this velocity, but cannot confirm the origin of these components. If associated with the Near 3 kpc feature, the clouds would be approximately 0.5 kpc below the disk.

\subsection{HD~163522}\label{app:hd163522_desc}
\begin{figure}[t!]
    \centering
    \includegraphics[width=\columnwidth]{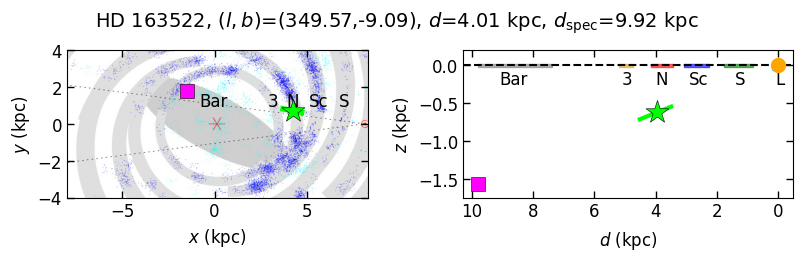}
    \includegraphics[width=\columnwidth]{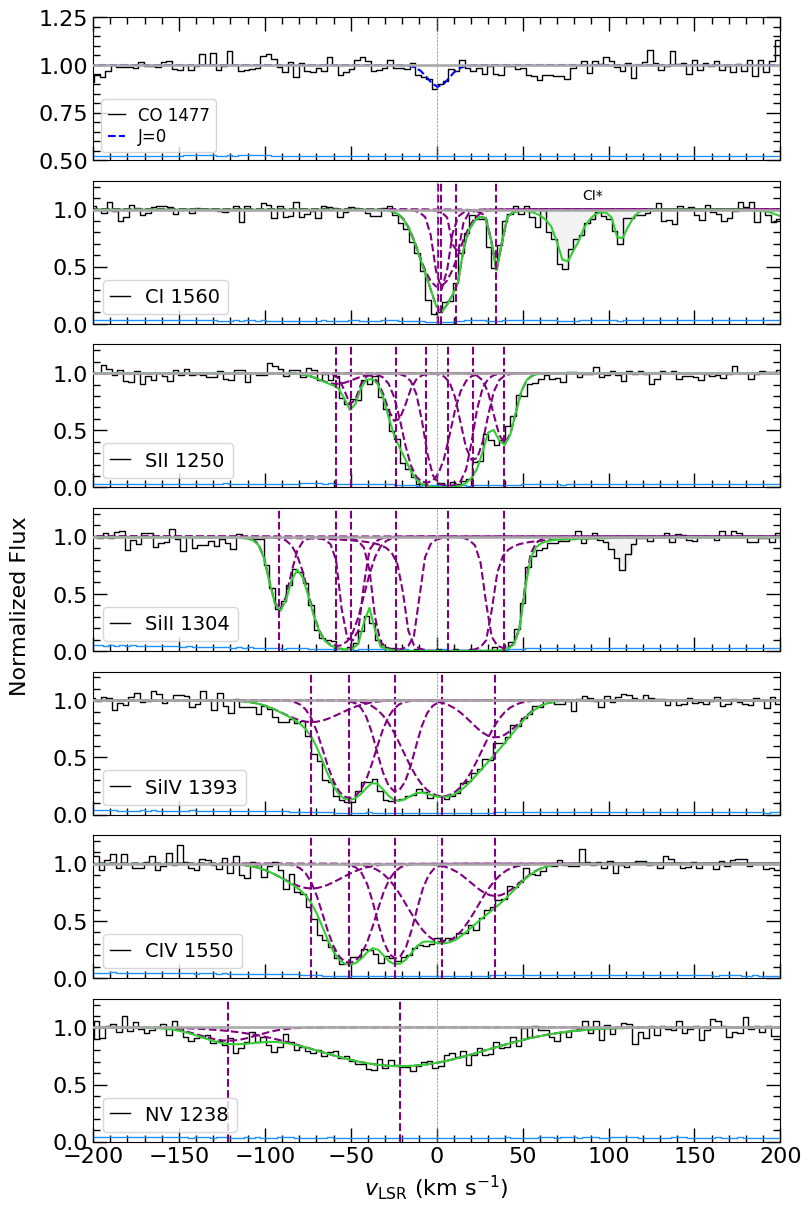}
    \caption{Same as for Figure \ref{fig:hd165955}, but for the HST--STIS E140M spectrum of HD 163522. The green star represents the Gaia DR3 distance $d=4.01^{+0.56}_{-0.47}$ kpc. The magenta square represents the spectroscopic distance $d=9.92$ kpc from \citet{jenkins2009}.}
    \label{fig:hd163522}
\end{figure}

\emph{HD 163522} has a Gaia DR3 distance $d=4.01^{+0.56}_{-0.47}$ kpc that is inconsistent with a higher spectroscopic distance $d=9.92$ kpc from \citet{jenkins2009}, see Figure \ref{fig:hd163522}. 
This Quadrant IV sight line was previously studied with FUSE by \citet{savage1990}, who calculate a similar spectroscopic distance $d=9.4$ kpc. 
At the higher spectroscopic distance, the star lies behind the GC and the region under all four inner arms is probed. 
At the higher end of the Gaia distance range, the star lies below the Norma arm. We observe the $J=0$ CO $\lambda$1477 transition.
Our model of differential Galactic rotation (see Section \ref{subsec:corotation}) gives a co-rotating velocity range of $-177 \leq v$ (\kms) $ \leq 0$.
We note that \citetalias{reid2019} observe gas associated with the Near-3-kpc arm at \vlsr = $-$108.8 \kms. We do observe \ion{N}{5} at that velocity. If somehow associated, this cloud would lie nearly 1 kpc below the Near 3 kpc arm.

\subsection{HD~148422}\label{app:hd148422_desc}
\begin{figure}[t!]
    \centering
    \includegraphics[width=\columnwidth]{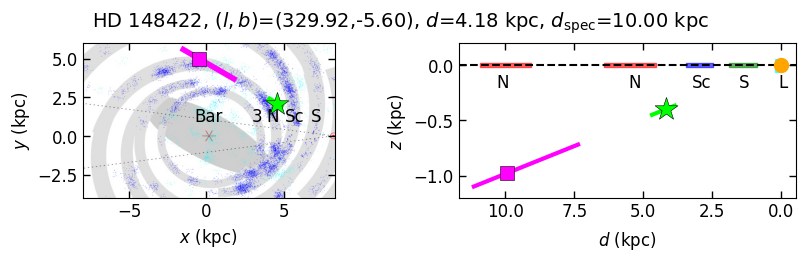}
    \includegraphics[width=\columnwidth]{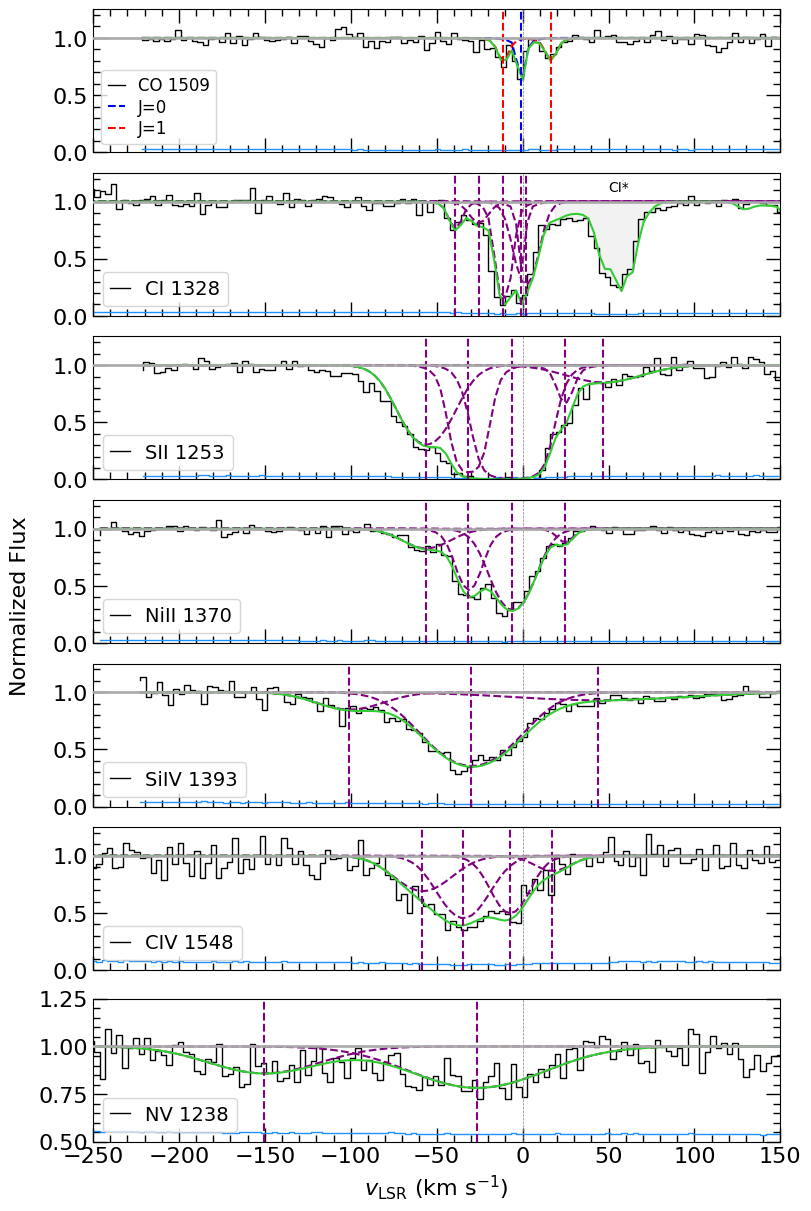}
    \caption{Same as for Figure \ref{fig:hd165955}, but for the HST--STIS E140M spectrum of HD 148422. The green star shows the Gaia DR3 distance of $d=4.18^{+0.52}_{-0.31}$ kpc. The magenta square shows the position of the star at the spectroscopic distance of $d=10.0^{+1.2}_{-2.6}$ kpc from \citet{bowen2008}. At this higher distance, the sight line would probe 0.5--1.0 kpc under an extended section along the Norma tangent, as indicated in the top left panel. }
    \label{fig:hd148422}
\end{figure}

\emph{HD 148422} has a spectroscopic distance of $d=10.0^{+1.2}_{-2.6}$ kpc from \cite{bowen2008} that is substantially higher than the reported Gaia DR3 distance of $d=4.18^{+0.52}_{-0.31}$ kpc, as shown in Figure \ref{fig:hd148422}. 
This star was included in an optical study by \citet{smartt1997} in an effort to discover the origination point for ejected blue supergiants.
At the lower Gaia distance, the sight line is expected to probe under the Carina and Crux-Centaurus arms in Quadrant IV. 
However, at the higher end of the spectroscopic distance range, the sight line may pass below an extended section of the Norma arm as a result of the sight line's large longitude offset from $l=0$ of approximately 30$^\circ$. 
Our model of differential Galactic rotation (see Section \ref{subsec:corotation}) gives a co-rotating velocity range of $-109 \leq v$ (\kms) $ \leq 75$.
We observe one component of the ground state $J=0,1$ molecular CO transitions at 1419, 1447, 1477, and 1509 {\AA} near 0 \kms.

\subsection{HD~164340}\label{app:hd164340_desc}
\begin{figure}[t!]
    \centering
    \includegraphics[width=\columnwidth]{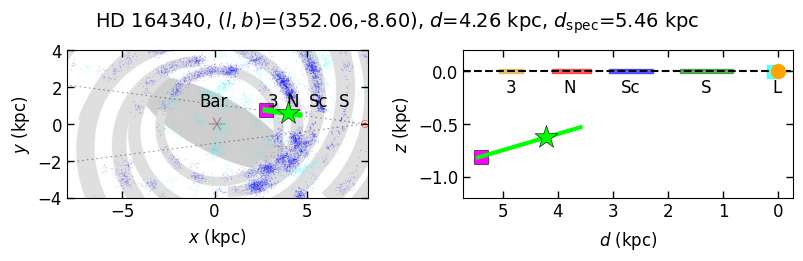}
    \includegraphics[width=\columnwidth]{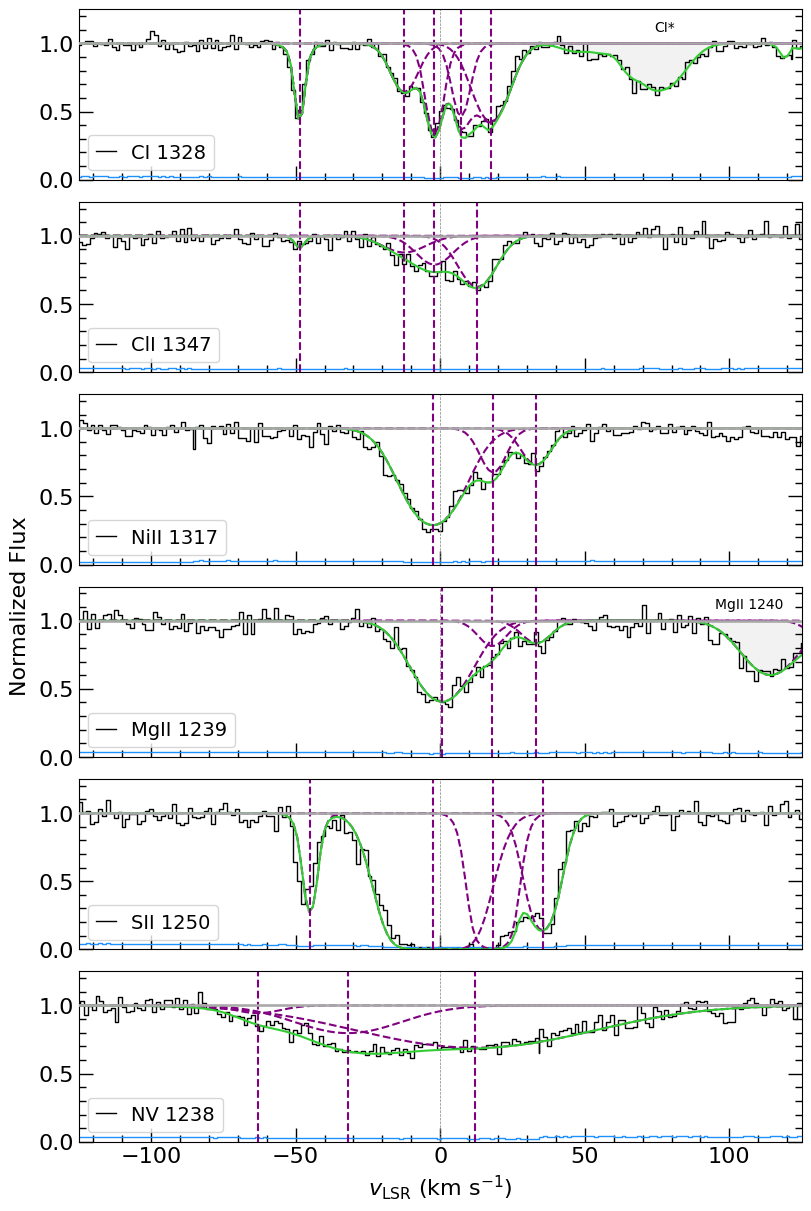}
    \caption{Same as for Figure \ref{fig:hd165955}, but for the HST--STIS E140H spectrum of HD 164340. The green star shows the Gaia DR3 distance of $d=4.26^{+1.25}_{-0.63}$ kpc. The magenta square shows the position of the star from the spectroscopic distance $d=5.46$ kpc of \citet{sembach1993}.}
    \label{fig:hd164340}
\end{figure}

\emph{HD 164340} has a Gaia DR3 distance $d=4.26^{+1.25}_{-0.63}$ whose upper end is consistent with the spectroscopic distance $d=5.46$ kpc from \citet{sembach1993}, see Figure \ref{fig:hd164340}. 
Under both distance regimes the region under all 4 inner arms may be probed in Quadrant IV. 
This sight line was observed with the STIS E140H grating and central wavelength setting at 1271 {\AA}. Since this short wavelength setting does not cover the stronger UV CO lines, we only have an upper limit on the ``strongest" CO transition ($J=0$; $\lambda$1344; $f=4.14\times10^{-3}$) covered by this spectrum. 
The only high ion covered by this STIS setting is \ion{N}{5}. 
Our model of differential Galactic rotation (see Section \ref{subsec:corotation}) gives a co-rotating velocity range of $-187 \leq v$ (\kms) $ \leq 0$.
\citetalias{reid2019} observe gas associated with the Near-3-kpc arm centered near $-$92.0 \kms. We do not observe UV gas at this velocity.

\subsection{HD~179407}\label{app:hd179407_desc}
\begin{figure}[t!]
    \centering
    \includegraphics[width=\columnwidth]{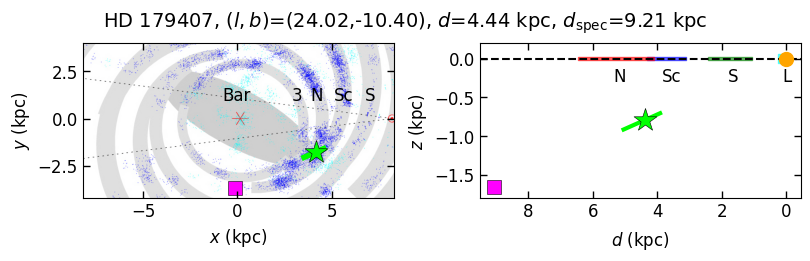}
    \includegraphics[width=\columnwidth]{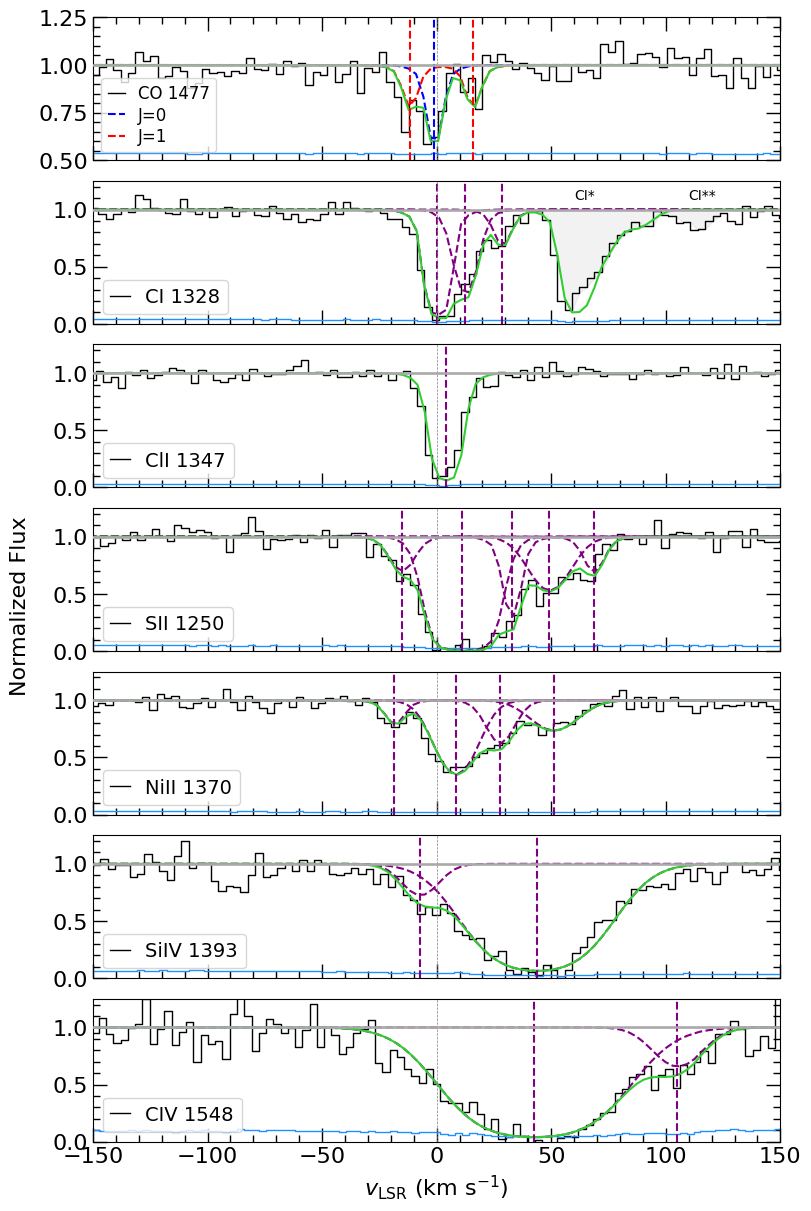}
    \caption{Same as for Figure \ref{fig:hd165955}, but for the HST--STIS E140M spectrum of HD 179407. The green star shows the Gaia DR3 distance of $d=4.44^{+0.69}_{-0.47}$ kpc. The magenta square shows the position of the star from the spectroscopic distance $d=9.21$ kpc of \citet{jenkins2009}. The sight line is of high enough longitude in Quadrant I that the sight may probe under an extended length of Scutum and Norma.}
    \label{fig:hd179407}
\end{figure}

\emph{HD 179407} has a Gaia DR3 distance $d=4.44^{+0.69}_{-0.47}$ kpc that is inconsistent with a higher spectroscopic distance $d=9.21$ kpc from \citet{jenkins2009}, see Figure \ref{fig:hd179407}.
At the highest end of the Gaia distance range, the star lies under the region between Scutum and Norma, and could be expected to probe under both Scutum and Sagittarius in Quadrant I. 
At the higher spectroscopic distance, the sight line would extend a few kiloparsecs under Scutum and 
may also skirt under the edge of the Norma arm. 
We observe the ground-state $J=0,1$ CO $\lambda$1477 transitions.
Our model of differential Galactic rotation (see Section \ref{subsec:corotation}) gives a co-rotating velocity range of $0 \leq v$ (\kms) $  \leq 128$.

\subsection{HD~167402}\label{app:hd167402_desc}
\begin{figure}[t!]
    \centering
    \includegraphics[width=\columnwidth]{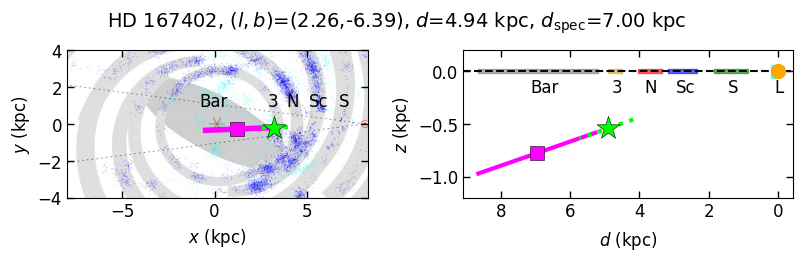}
    \includegraphics[width=\columnwidth]{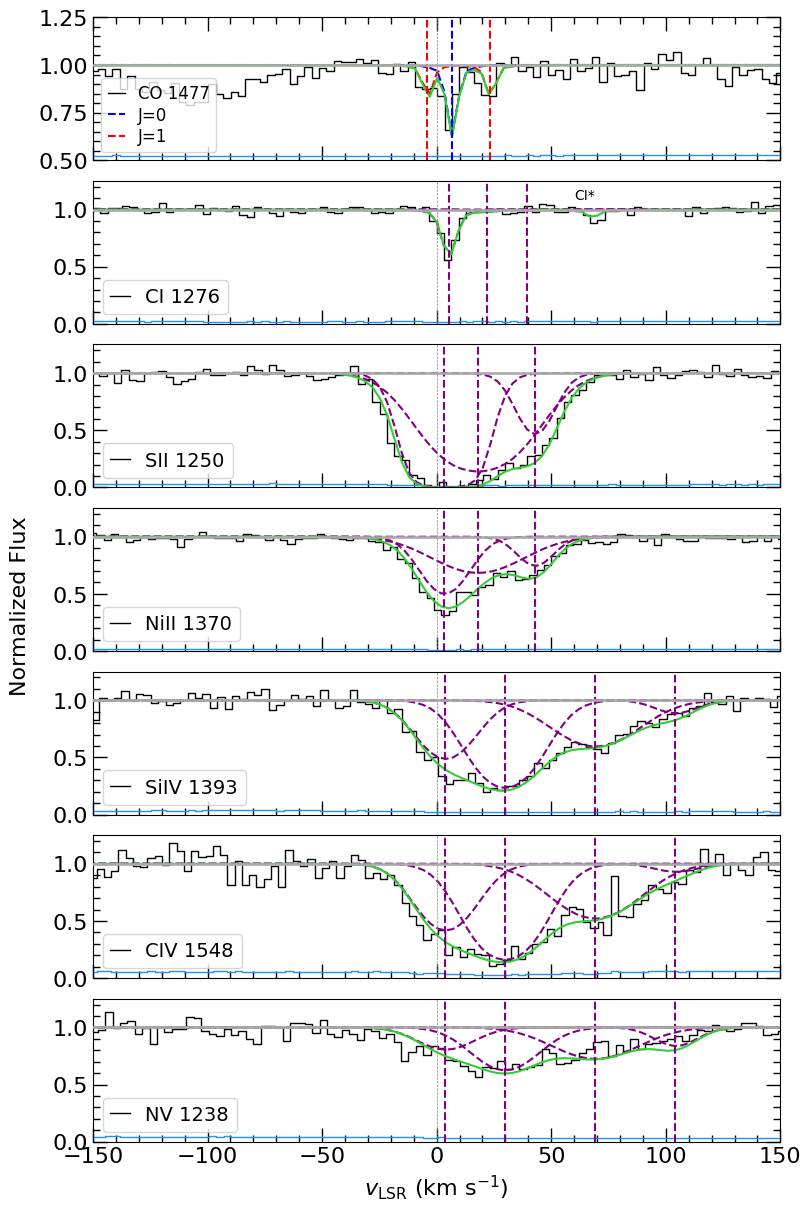}
    \caption{Same as for Figure \ref{fig:hd165955}, but for the HST--STIS E140M spectrum of HD 167402. The green star shows the Gaia DR3 distance of $d=4.94^{+0.83}_{-0.73}$ kpc. The magenta square shows the position of the star from the spectroscopic distance $d=7.0\pm1.7$ kpc of \citet{savage2017}.}
    \label{fig:hd167402}
\end{figure}

\emph{HD 167402} is the lowest-longitude sight line in our sample. The Gaia DR3 distance of 4.94$^{+0.83}_{-0.73}$ kpc overlaps with the low end of the spectroscopic distance of $7.0\pm1.7$ kpc, as shown in Figure \ref{fig:hd167402}. 
In this distance regime, the region under all four arms is expected to be probed in Quadrant I. 
Our model of differential Galactic rotation (see Section \ref{subsec:corotation}) gives a co-rotating velocity range of $-6 \leq v$ (\kms) $ \leq 129$.
We observe the ground-state $J=0,1$ molecular CO transitions at 1419, 1447, 1477, and 1509 {\AA} with a velocity centroid at 6.5 \kms.
There are multiple components at positive velocities that are outside the range of velocities for co-rotation with any arm. This sight line and its intermediate and high-velocity clouds are discussed at length in \citet{savage2017} and \citet{cashman2021}.
We note that \citetalias{reid2019} associate gas near $-$43.6 \kms\ with the Near 3 kpc arm. We do see gas at this velocity, but cannot confirm a Near 3 kpc origin. If somehow associated, it would lie approximately 0.5 kpc below the midplane.

\subsection{HD~156359}\label{app:hd156359_desc}
\begin{figure}[t!]
    \centering
    \includegraphics[width=\columnwidth]{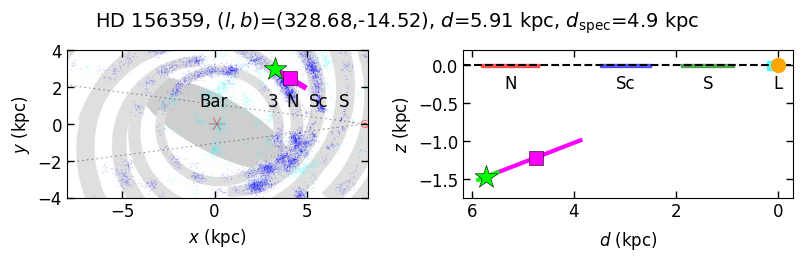}
    \includegraphics[width=\columnwidth]{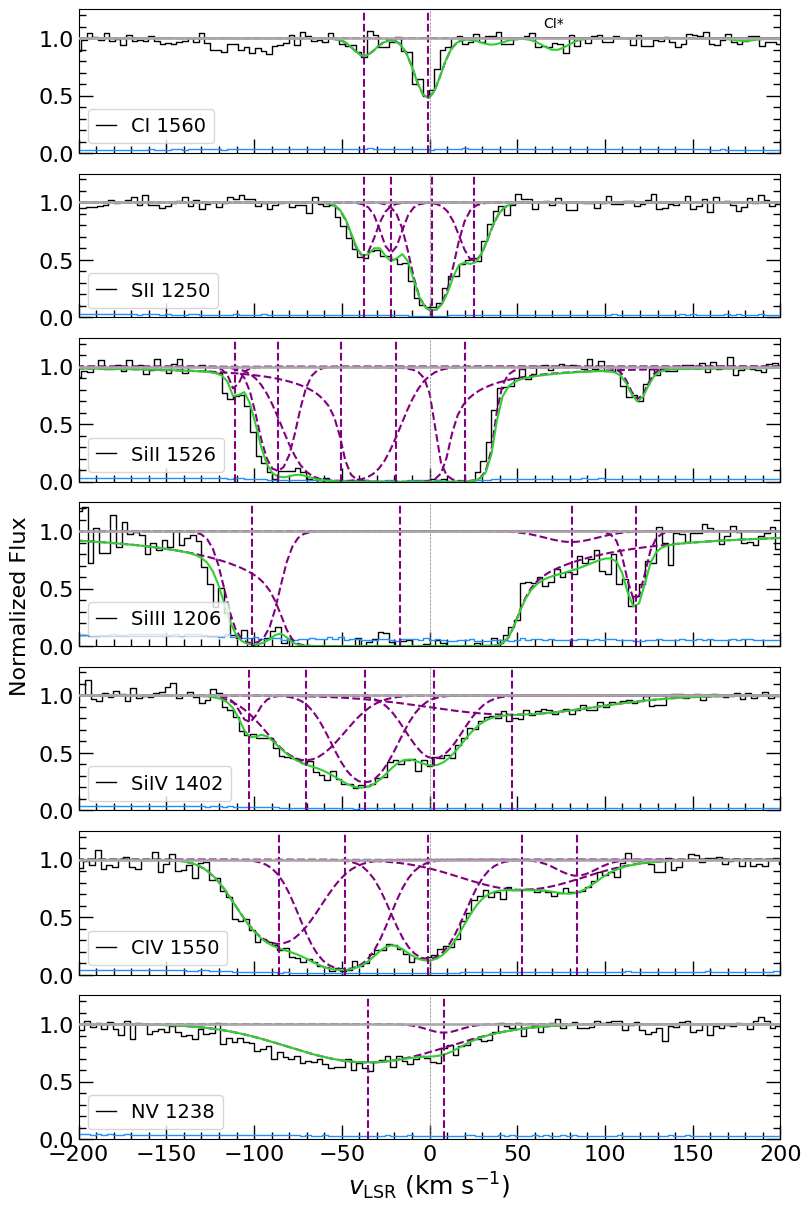}
    \caption{Same as for Figure \ref{fig:hd165955}, but for the HST--STIS E140M spectrum of HD 156359. The green star shows the Gaia DR3 distance of $d=5.91^{+0.16}_{-0.23}$ kpc. The magenta square shows the position of the star with the spectroscopic distance $d=4.9^{+1.1}_{-0.9}$ kpc from J. Ma\'{i}z-Apell\'{a}niz (priv. comm.).}
    \label{fig:hd156359}
\end{figure}

\emph{HD 156359} is a high-latitude runaway O star whose sight line has been previously studied by \citet{sembach1991} with the International Ultraviolet Explorer (IUE) and in \citet{cashman2023} using the Far-Ultraviolet Spectroscopic Explorer (FUSE) and HST/STIS. 
The higher end of the spectroscopic distance $d=4.9^{+1.1}_{-0.9}$ kpc from J. Ma\'{i}z-Apell\'{a}niz (priv. comm.) overlaps with the Gaia DR3 distance of $d=5.91^{+0.16}_{-0.23}$ kpc, see Figure \ref{fig:hd156359}. 
The Quadrant IV star is at the highest latitude ($\sim$15$^\circ$), as well as the largest longitude offset in our entire sample, $\sim$31$^\circ$ from $l=0$. 
Our model of differential Galactic rotation (see Section \ref{subsec:corotation}) gives a co-rotating velocity range of $-102 \leq v$ (\kms) $ \leq 0$.
At the highest possible distance, the sight line would probe under the Carina, Crux-Centaurus, and Norma arms.  
Although many UV CO transitions are covered by the STIS E140M spectrum, we observe no CO along this line of sight. This may be due to the high latitude of the sight line. 
Thus, this non-detection could imply a limit to the height of the UV CO in the disk in this direction to a few tens of parsecs. 
We observe only one neutral gas species, \ion{C}{1}, with only one component centered near 0 \kms.
There are intermediate velocity cloud components observed in the low and high ions near $+50$ and $+80$ \kms. The higher velocity components observed near $+100$ \kms\ are known to be associated with the Complex WE high-velocity cloud and are discussed in \citet{cashman2023}.

\begin{figure}[t!]
    \centering
    \includegraphics[width=\columnwidth]{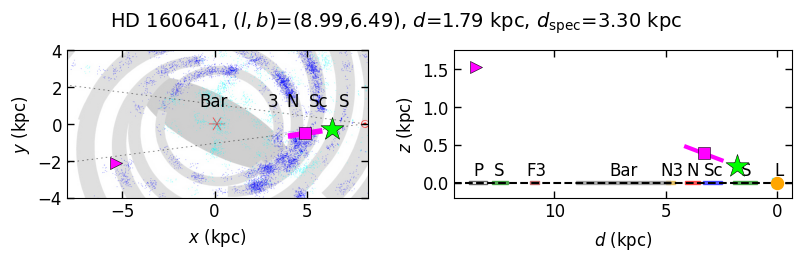}
    \includegraphics[width=\columnwidth]{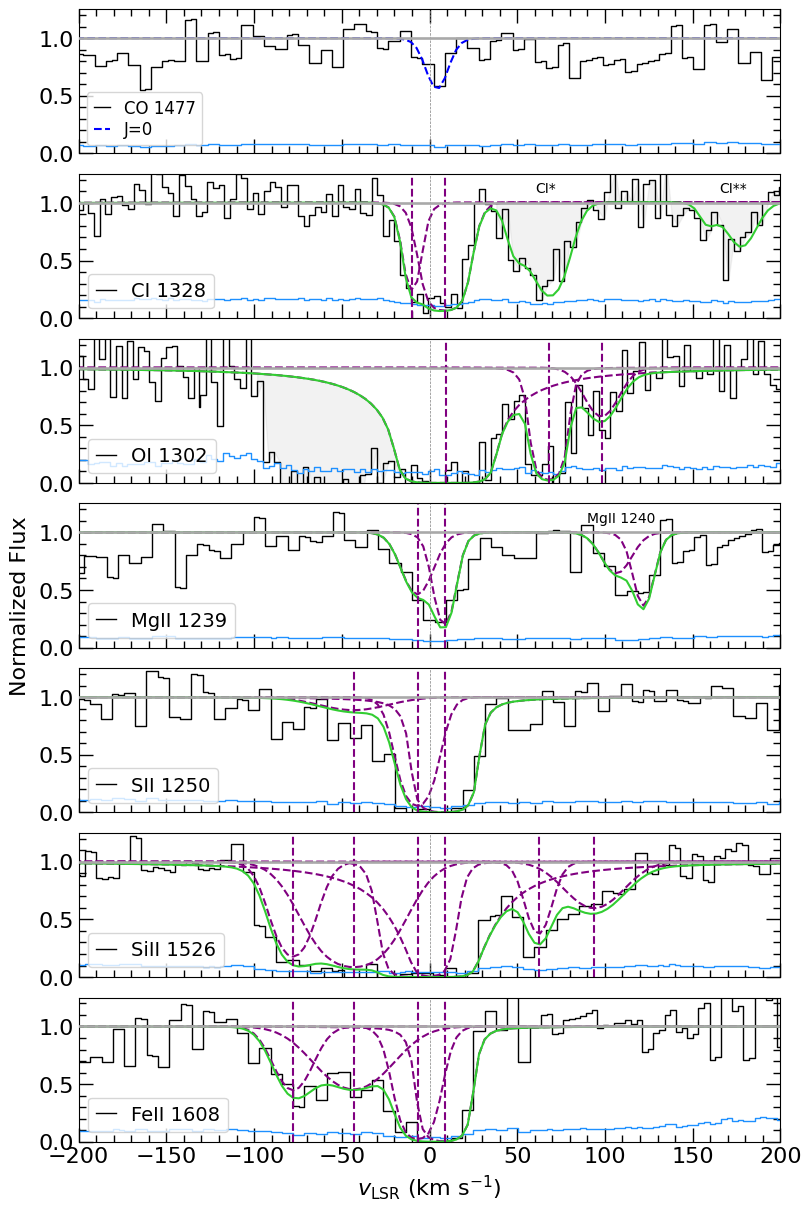}
    \caption{Same as for Figure \ref{fig:hd165955}, but for the HST--STIS E140M spectrum of HD 160641. The green star shows the Gaia DR3 distance $d=1.79^{+0.09}_{-0.10}$ kpc. The magenta square shows $d=3.3\pm0.8$ kpc from a recent reclassification to a hot subdwarf sdOC9.5II-III:He40 \citep{bobylev2015,drilling2013,lynas1987}. The right-pointing green triangle in the top two panels shows spectroscopic distance $d=13.6$ kpc from \citet{reed2003} for a O9.5IA-B star. At this distance, the sight line would additionally probe over 1 kpc above the Sagittarius, Scutum, and the Near-3-kpc arms on the opposite side of the GC.
    }
    \label{fig:hd160641}
\end{figure}

\begin{figure}[t!]
    \centering
    \includegraphics[width=\columnwidth]{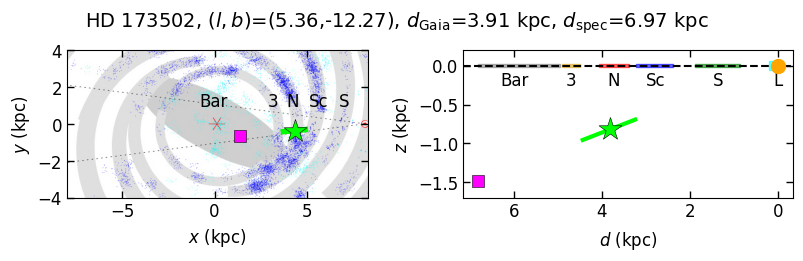}
    \includegraphics[width=\columnwidth]{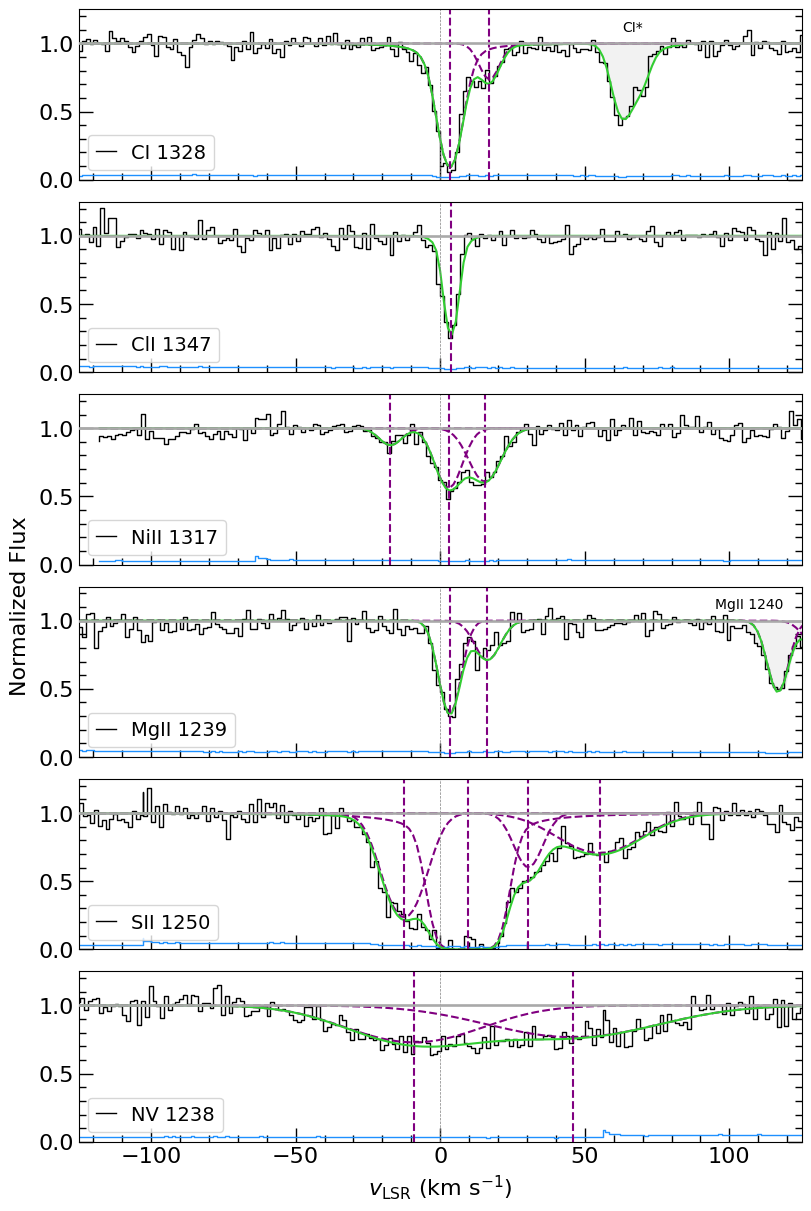}
    \caption{Same as for Figure \ref{fig:hd165955}, but for the HST--STIS E140H spectrum of HD 173502. The green star represents the Gaia DR3 distance $d=3.91^{+0.63}_{-0.58}$ kpc. The magenta square represents the spectroscopic distance $d=6.970$ kpc from \citet{sembach1993}.}
    \label{fig:hd173502}
\end{figure}

\clearpage
\bibliography{ref}{}
\bibliographystyle{aasjournal}

\end{document}